\documentclass[prx,amsmath,amssymb, notitlepage, twocolumn,
nofootinbib,
superscriptaddress,
longbibliography
]{revtex4-1}

\usepackage{amsmath}
\usepackage{tabularx,graphicx}
\usepackage{epstopdf}
\usepackage{graphicx}
\usepackage{latexsym}
\usepackage{amssymb}
\usepackage{amsmath}
\usepackage{color, colortbl}
\usepackage{psfrag}
\usepackage{bbm}
\usepackage{bm}
\usepackage{titlesec}
\usepackage{dsfont}
\usepackage{feynmp}
\usepackage{slashed}
\usepackage{multirow}
\newcommand{\mb}[1]{\mathbf{#1}}
\newcommand{\mc}[1]{\mathcal{#1}}

\newcommand{\beq}{\begin{eqnarray}}
\newcommand{\eeq}{\end{eqnarray}}

\newcommand{\la}{\langle}
\newcommand{\ra}{\rangle}

\newcommand{\Tr}{{\rm Tr}}

\newcommand{\bsp}{\begin{split}}
\newcommand{\esp}{\end{split}}

\newcommand{\ie}{{\it i.e., }}
\newcommand{\eg}{{\it e.g., }}
\newcommand{\gcs}{{\rm{CS}}_g}

\usepackage{color}
\definecolor{darkblue}{rgb}{0.,0.,0.4}
\definecolor{darkred}{rgb}{0.5,0.,0.}
\definecolor{BlueViolet}{RGB}{138,43,226}
\definecolor{SkyBlue}{RGB}{30,144,255}
\definecolor{DarkGreen}{RGB}{0,100,0}
\usepackage[pdftex,colorlinks=true,linkcolor=darkblue,citecolor=blue,urlcolor=darkred]{hyperref}
\usepackage[normalem]{ulem}

\renewcommand{\vec}[1]{\bm{#1}}

\begin{document}

\title{Field-induced QCD$_3$-Chern-Simons quantum criticalities in Kitaev materials}
\author{Liujun Zou}
\affiliation{Department of Physics, Harvard University, Cambridge, MA 02138, USA}
\affiliation{Department of Physics, Massachusetts Institute of Technology, Cambridge, MA 02139, USA}

\author{Yin-Chen He}
\affiliation{Perimeter Institute for Theoretical Physics, Waterloo, Ontario N2L 2Y5, Canada}

\begin{abstract}

Kitaev materials are promising for realizing exotic quantum spin liquid phases, such as a non-Abelian chiral spin liquid. 
Motivated by recent experiments in these materials, we theoretically study the novel field-induced quantum phase transitions from the non-Abelian chiral spin liquid to the symmetry-broken zigzag phase and to the trivial polarized state. Utilizing the recently developed dualities of gauge theories, we find these transitions can be described by critical bosons or gapless fermions coupled to emergent non-Abelian gauge fields, and the critical theories are of the type of a QCD$_3$-Chern-Simons theory. We propose that all these exotic quantum phase transitions can potentially be direct and continuous in Kitaev materials, and we present sound evidence for this proposal. 
Therefore, besides being systems with intriguing quantum magnetism, Kitaev materials may also serve as table-top experimental platforms to study the interesting dynamics of emergent strongly interacting quarks and gluons in $2+1$ dimensions. 
\end{abstract}

\maketitle

\tableofcontents

\section{Introduction} \label{sec: intro}

Understanding the universal properties of quantum phases and quantum phase transitions is one of the central goals of physics. Both quantum phases and phase transitions may be characterized by interesting universality classes, but perhaps partly due to the facts that quantum phases are often much simpler to understand and that they occupy the most regions of the phase diagram, more efforts have been devoted to exploring exotic quantum phases, rather than phase transitions.

However, the understanding of a quantum phase transition between two phases necessarily involves the understanding of the intricate interplay among all the degrees of freedom in each phase, so such understanding offers not only the understanding of the possible universality class of the transition itself, but also unified understanding of the nearby phases. Therefore, in certain sense understanding quantum phase transitions is of more fundamental importance \cite{Sachdev_book2011}.

The best understood examples of quantum phase transitions are between a symmetric phase and a spontaneously-symmetry-broken (SSB) phase. The critical theory for such a phase transition is often formulated in terms of some fluctuating local order parameters, and the associated universal critical physics can be obtained by a renormalization group (RG) treatment of this critical theory. This is known as the Landau-Ginzburg-Wilson paradigm \cite{Sachdev_book2011}. In recent years, exploring exotic quantum phase transitions beyond this conventional paradigm has become a frontier of condensed matter physics. 
For an incomplete list of these studies, see Refs. \cite{deccp, deccplong, Senthil2008, Maissam2014_FQHtransition, Tarun2013_QHtransition, LuLee, Zou2016, Tsui2017, Wang_DQCP_duality, You2017, You2017a, Lee2018_FCI, Gazit2018, BS2018}. However, many of these studies are mostly theoretical and relatively far from experiments, so it is important to search for exotic quantum phase transitions that are relevant to current experiments.

One occasion where an exotic quantum phase transition may occur is around a quantum spin liquid (QSL), \ie a spin system whose ground state exhibits nontrivial patterns of quantum entanglement \cite{Wen2004Book, QSLreview2019}.
Recently there has been exciting progress in material realizations of quantum spin liquids~\cite{QSLreview2019}, and an interesting class of materials is Kitaev materials (For recent reviews, see Refs. \cite{Kee2016, Hermanns_review, Trebst_Review, Winter_Review}). 
These Kitaev materials are believed to be described by Hamiltonians that are close to the Kitaev honeycomb model~\cite{Kitaev2006}, which is an exactly solvable model with three different types of QSL ground states (depending on parameters of Hamiltonian): a gapless QSL, a gapped Abelian QSL, and a gapped non-Abelian QSL. 

The gapped non-Abelian QSL, more precisely speaking, realizes an Ising topological order (ITO), which has two types of fractionalized excitations: a non-Abelian anyon $\sigma$ and a Majorana fermion. This state can be viewed as a $p+ip$ superconductor where the Bogoliubov quasi-particles therein are coupled to a dynamical $Z_2$ gauge field, and $\sigma$ plays the role of the $\pi$ flux in this superconductor. Just as a $p+ip$ superconductor, the ITO must break time reversal and mirror symmetries. In fact, it has a chiral edge mode and is supposed to have a quantized thermal Hall conductance $\kappa_{xy}=1/2$ in units of $(\pi/6)(k_B^2T/\hbar)$. This property offers an experimentally feasible method to detect the ITO.

\begin{figure}
    \centering
    \includegraphics[width=\linewidth]{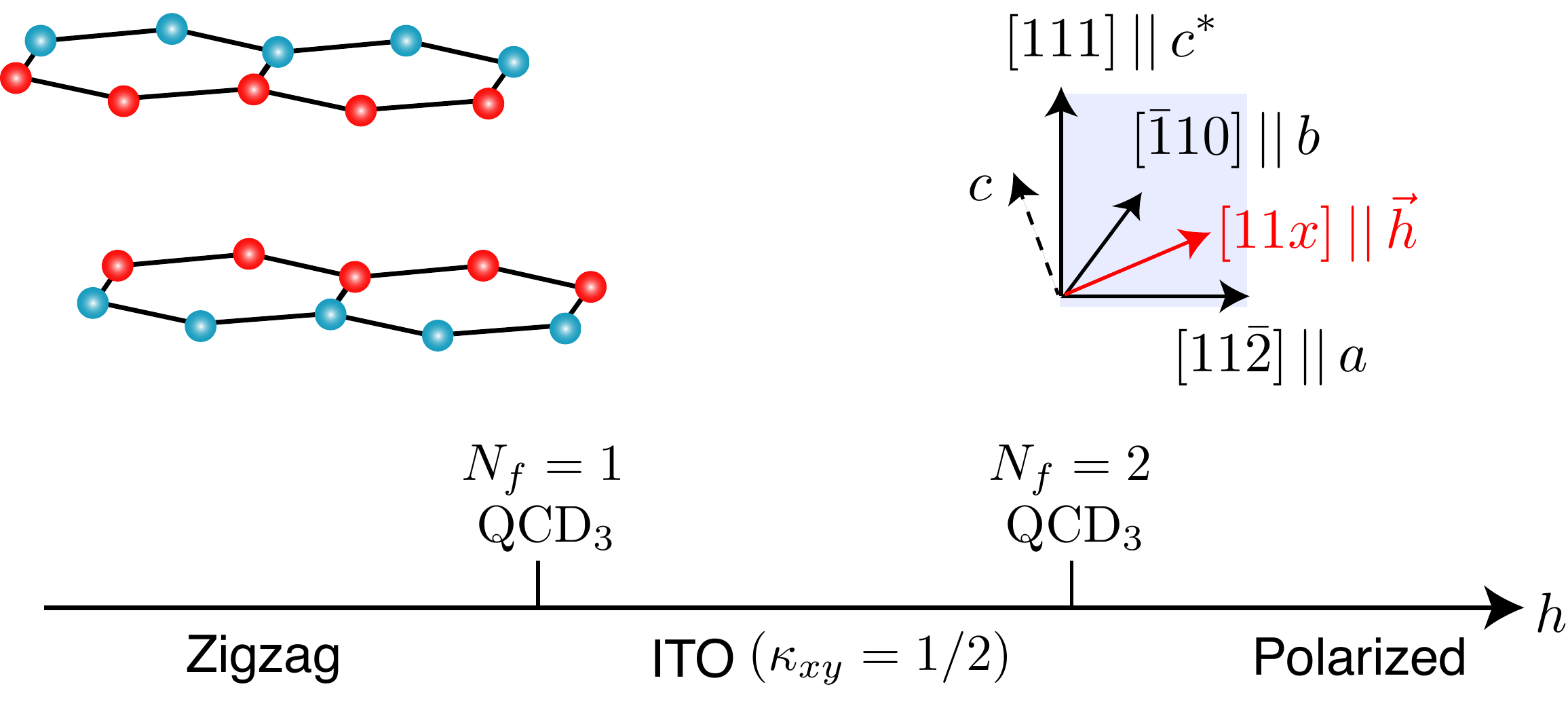}
    \caption{Under an external magnetic field, a Kitaev material may go through different phases: a zigzag ordered state, a non-Abelian chiral quantum spin liquid with Ising topological order (ITO), and a trivial polarized state.
    A similar phase diagram was observed in experimental~\cite{Rucl3_kappaxy} and numerical work~\cite{Kee2019}.
    The two phase transitions are described by two different QCD$_3$-Chern-Simons theories, which have emergent gapless Dirac fermions (with different fermion flavor numbers $N_f=1,2$) coupled to a $U(2)$ Chern-Simons gauge field.
    These two QCD$_3$-Chern-Simons transitions require a high symmetry of the system to be stable, and it can be satisfied if the magnetic field is on the $ac^*$ plane of Kitaev materials (\eg $\alpha$-RuCl$_3$).
    With certain details modified, the $N_f=1$ QCD$_3$ theory can also describe the phase transition from the ITO to other magnetic ordered states in other Kitaev materials, such as the Neel and stripy phases~\cite{Kee2016, Hermanns_review, Trebst_Review, Winter_Review}.
    }
    \label{fig:summary}
\end{figure}

Among the various Kitaev materials, $\alpha$-RuCl$_3$ has received significant attention recently. It is found that the ground state of $\alpha$-RuCl$_3$ is magnetically ordered in the absence of an external magnetic field \cite{Fletcher1967, Kobayashi1992}. Specifically, the spins are ordered in a zigzag pattern \cite{Sears2015, Coldea2015, Cao2016}, as shown in Fig.~\ref{fig:summary}.
The Heisenberg and $\Gamma$ interactions (introduced in Eq.~\eqref{eq:Ham}) are believed to be responsible for the zigzag order \cite{Chaloupka2013_zigzag,Jeffrey2014}, although the signs and strengths of these interactions in the real material are not fully settled down. 
It is also suggested that the $\Gamma$ term may help to stabilize a QSL~\cite{Gohlke_Gamma}. 
Notice a similar zigzag order has also been found in another Kitaev material, Na$_2$IrO$_3$ \cite{Hill2011, Taylor2012, Cao2012}.
Upon applying an external magnetic field, the zigzag magnetic order in $\alpha$-RuCl$_3$ melts \cite{banerjee2017neutron, Lampen-Kelley2018,Jansa2018, Banerjee2018}. If the magnetic field is strong enough, the system will become a trivial polarized state. 
Remarkably, the measured thermal Hall conductance in certain range of field strengths is quantized exactly at $\kappa_{xy}=1/2$~\cite{Rucl3_kappaxy}, which strongly suggests that an ITO is induced by the Zeeman field. 
Although there is some subtlety in interpreting this experiment \cite{Ye2018,Rosch2018,Cookmeyer2018}, and the results therein need to be confirmed by further studies, this discovery has triggered great excitement.

These experimental results suggest that $\alpha$-RuCl$_3$ exhibits only three phases upon increasing the external magnetic field, namely, the zigzag order, the ITO, and the trivially polarized state~\cite{Rucl3_kappaxy}.
A natural question immediately arises: What is the nature of the two phase transitions (zigzag-ITO and ITO-polarized state) as the magnetic field is tuned?
In this paper, we manage to tackle this problem theoretically. 
Intriguingly, we find that such quantum phase transitions are strikingly different from the conventional phase transitions, owing to the emergence of some deconfined non-Abelian gauge fields.
In particular, these quantum critical points mimic the QCD theories in 2+1 dimensions, which have emergent quarks and gluons that are strongly interacting with each other.
Moreover, these critical theories have interesting duality properties, namely, they can  be described either by critical bosons interacting with a $U(2)$ Chern-Simons gauge field, or by gapless Dirac fermions interacting with a $U(2)$ Chern-Simons gauge field.
In recent years, dualities of interacting gauge theories have generated huge theoretical enthusiasm in both the condensed matter and the high energy communities~\cite{sonphcfl,wangsenthil15b,MaxAshvin15,Seiberg2016395,karchtong,Hsin2016}, and the Kitaev materials may be one of the few experimental platforms~\cite{Lee2018_FCI} to study theories that have interesting duality properties.

We remark that our discussion on these QCD$_3$-Chern-Simons quantum criticalities is very general, and it only relies on the symmetries of the Kitaev materials, but not their microscopic details (\eg the precise spin Hamiltonian that describes the Kitaev material). Interestingly, some numerical evidence of these quantum phase transitions has been found recently \cite{Kee2019}. So it is relevant and timely to study such transitions more thoroughly.

Our results are schematically summarized in Fig.~\ref{fig:summary}, and the rest of the paper is organized as follows. 
In Sec. \ref{sec: symmetries}, we first review the global symmetries of some representative Kitaev materials, including $\alpha$-RuCl$_3$, Na$_2$IrO$_3$, etc. 
Based on these general symmetry properties, in Sec. \ref{sec: QCD3}, we discuss the quantum phase transitions from the ITO state to the zigzag phase and to the trivial polarized state, and we find they are described by emergent QCD$_3$-Chern-Simons gauge theories. 
In Sec. \ref{sec: exp}, we discuss the experimental signatures of the QCD$_3$-Chern-Simons quantum critical points. Finally, we summarize our results and discuss some future directions in Sec. \ref{sec: discussion}. The appendices contain various technical details, some of which present powerful methods that can be adopted to study various related problems.

\section{Symmetries of the materials and models} \label{sec: symmetries}

\begin{figure}
	\centering
	\includegraphics[width=0.7\linewidth]{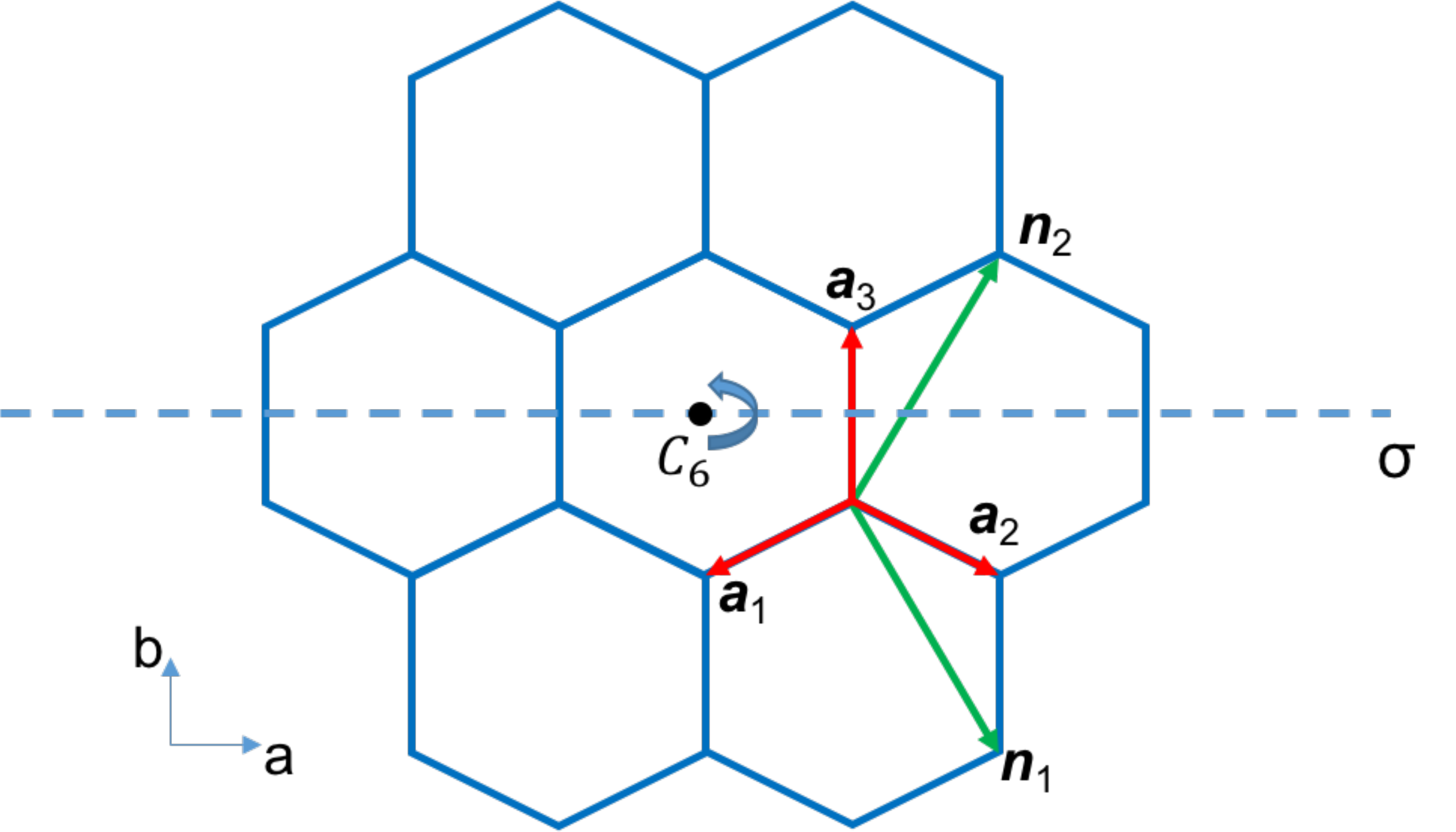}
	\caption{The honeycomb lattice. The $x$-bond, $y$-bond and $z$-bond are along the direction of $\vec a_1$, $\vec a_2$ and $\vec a_3$, respectively. The $a$- and $b$-axes are shown in the figure, and the $c^*$-axis is perpendicular to the paper and pointing outwards.}
	\label{fig:honeycomb-lattice}
\end{figure}

For concreteness, we start by introducing the Hamiltonian of a Kitaev material, and we stress again that it is only the global symmetries rather than the detailed Hamiltonian that play a role in the following discussion on the novel quantum phase transitions. A general Hamiltonian for a Kitaev material under a magnetic field up to nearest-neighbor coupling can be written as
\begin{align}
\label{eq:Ham}
H 
=&  \sum_{\langle i j\rangle \in \alpha} K_\alpha S_i^\alpha S_j^\alpha - \sum_{i} \vec h\cdot\vec S_{i} +J_H \sum_{\langle i j\rangle \in \alpha} \vec S_i \cdot \vec S_j  \nonumber \\ 
&+  \sum_{\langle i j\rangle \in \alpha} \Gamma_\alpha S_i^\beta S_j^\gamma +\cdots \end{align}
where the second and third terms are the familiar Zeeman field term (with the $g$-tensor suppressed) and Heisenberg interaction, respectively. 
The $K_\alpha S_i^\alpha S_j^\alpha$ and $\Gamma_\alpha S_i^\beta S_j^\gamma$ terms are often referred to as the Kitaev term and $\Gamma$ term, respectively.
The $\cdots$ term includes all other symmetry-allowed terms, such as the $\Gamma'$ term (defined below), which was argued to be important~\cite{Kee2019} for realizing the field-induced experimental phase diagram~\cite{Rucl3_kappaxy} (see Fig.~\ref{fig:summary}).
On the bond $\langle ij\rangle_x$, the $x$-bond connecting site $i$ and site $j$, the $K$ and $\Gamma$ terms read $K_x S_i^x S_j^x$ and $\Gamma_x (S_i^y S_j^z+S_i^z S_j^y)$, respectively. As for the $\Gamma'$-term, on the bond $\langle ij\rangle_x$, it reads $\Gamma_x' (S_i^x S_j^y + S_i^y S_j^x + S_i^x S_j^z + S_i^z S_j^x)$. 
Similar notation is used for the $y$- and $z$-bonds. We denote the field direction $\vec h =(h_x, h_y, h_z)$ as  $[h_xh_yh_z]$, and the field direction $\vec h =(-h_x, h_y, h_z)$ as  $[\bar h_xh_yh_z]$, etc.

This Hamiltonian is exactly solvable if only the Kitaev term is present. In this case, it hosts two spin liquid ground states.
In particular, if $|K_x| < |K_y| +|K_z|$, etc., the ground state is a gapless $Z_2$ QSL with two Majorana cones.
Under a small magnetic field, as shown by Kitaev, the Hamiltonian gives rise to an ITO, a non-Abelian chiral QSL ground state. Under a very large magnetic field, the spins are trivially polarized \cite{Kitaev2006}. 

Once perturbed away from the pure Kitaev model, the Hamiltonian is no longer exactly solvable. 
It is still an open issue about the precise values of the coupling constants of each interaction term in Kitaev materials. In this paper, we will be primarily concerned with the symmetries of the system, and will not worry about the microscopic interaction strengths.

Some representative Kitaev materials, including $\alpha$-RuCl$_3$, Na$_2$IrO$_3$, etc., are layered quasi-two-dimensional materials with point group symmetry $C_2/m$ \cite{Coldea2015, Cao2016, Mai2019}\footnote{ We note that there is debate on the precise low-temperature symmetry of $\alpha$-RuCl$_3$, and some recent papers claim the symmetry should be $R\bar{3}$ \cite{Park2016, Do2017}. Here we assume the symmetry is C2/m, and the general method presented in this paper can also be straight- forwardly adopted to the case with $R\bar{3}$ symmetry.}.
The $C_2/m$ symmetry constrains that $K_x=K_y$, $\Gamma_x=\Gamma_y$, $\Gamma_x'=\Gamma_y'$.
It is expected that $K_z\sim K_{x,y}$ and $\Gamma_z\sim \Gamma_{x,y}$, $\Gamma_z'\sim \Gamma_{x,y}'$, but it is not crucial for our discussion.
Without the Zeeman field, the Hamiltonian enjoys the translation symmetry $T_{1,2}$ along $\vec n_{1, 2}$, inversion symmetry $C_2$, pseudo-mirror symmetry $\sigma^*$ (with the mirror axis perpendicular to the $z$ bond, or, equivalently, along the dashed line in Fig. \ref{fig:honeycomb-lattice}), and time-reversal symmetry $\mathcal{T}$ ($\vec S \rightarrow - \vec S$).
The pseudo-mirror symmetry $\sigma^*$ is the conventional mirror symmetry followed by a spin rotation symmetry $e^{i\pi S^y}e^{i\pi/2 S^z}$, 
\beq \label{eq: action of sigma*}
\begin{split}
\sigma^*: \quad & S_{\vec r}^x \rightarrow -S_{\sigma \vec r}^y, \\
 & S_{\vec r}^y \rightarrow -S_{\sigma \vec r}^x, \\
  & S_{\vec r}^z \rightarrow -S_{\sigma \vec r}^z.
\end{split}
\eeq
We remark that the spin flip symmetries $e^{i\pi S^\alpha}$ are broken due to the $\Gamma$ (or $\Gamma'$) terms.

Under a finite Zeeman field, the time reversal symmetry will be broken.
The pseudo-mirror symmetry  is also broken by a field along a generic direction.
There are special directions along which the pseudo-mirror symmetry or its combination with the time-reversal symmetry is preserved.
The details are summarized in Table~\ref{tab:symmetry}.
\begin{table}
\setlength{\tabcolsep}{0.2cm}
\renewcommand{\arraystretch}{1.4}
    \centering
    \begin{tabular}{c|ccccc} 
    \hline \hline
    & $T_{1,2}$ & $C_2$ & $\mathcal T$ & $\sigma^*$ & $\mathcal T \sigma^*$ \\ \hline
        $\vec h=0$ & Yes & Yes & Yes & Yes & Yes \\ 
        $\vec h \parallel [11x] $, in $ac^*$ plane & Yes & Yes & No & No & Yes \\
       $\vec h  \parallel  [\bar 110] $, parallel to $b$   &  Yes & Yes & No & Yes & No \\ 
       $\vec h \nparallel  [\bar 110],[11x]$ &  Yes & Yes & No & No & No \\
       \hline \hline

    \end{tabular}
    \caption{Symmetries of some representative Kitaev materials (including $\alpha$-RuCl$_3$, Na$_2$IrO$_3$, etc.) under the Zeeman field along different field directions $\vec h = (h_x, h_y, h_z)$.}
    \label{tab:symmetry}
\end{table}

The time reversal symmetry $\mathcal T$ or the pseudo-mirror symmetry $\sigma^*$ forbids a finite thermal Hall conductance.
Therefore, if the Zeeman field is parallel to the $b$ axis ($[\bar 1 10]$ direction), one cannot have an ITO unless $\sigma^*$ is spontaneously breaking.
On the other hand, if the Zeeman field is on the $ac^*$ plane, as is done in the thermal Hall experiments~\cite{Rucl3_kappaxy}, there is no symmetry that forbids the ITO.
However, this does not mean that we should expect an ITO for a field in a generic direction on the $ac^*$ plane.
After all, if one rotates the field on the $ac^*$ plane, there should be a phase transition between the ITO and its time-reversal partner.
This transition can be direct and continuous, or there can be an intermediate phase, \eg a $Z_2$ toric code phase, as one rotates the Zeeman field on the $ac^*$ plane.
In this paper, we will not pursue this direction.

Interestingly, the combination of time-reversal symmetry and pseudo-mirror symmetry, $\mathcal{T} \sigma^*$, is preserved for the field on the $ac^*$ plane ($\vec h \parallel[11x]$). This symmetry is crucial for the stability of the QCD$_3$-Chern-Simons quantum critical points, as we will discuss in Sec.~\ref{sec: QCD3}. We also note that Refs. \cite{Cao2016,Chun2015} reported that the zigzag order is on the $ac^*$ plane, which means this order preserves $\sigma^*$ but spontaneously breaks $\mc{T}\sigma^*$.

\section{QCD$_3$-Chern-Simons quantum criticalities} \label{sec: QCD3}

Utilizing the symmetry properties of the Kitaev materials discussed above, in this section, we study the possible exotic quantum phase transitions from the ITO state to the zigzag phase and to the polarized phase. These two latter states have no topological order, so these transitions can be viewed as confinement transitions of the ITO. 

At first glance, such confinement transitions are rather nontrivial if they can be continuous. 
To appreciate this, first notice the anyonic excitations in the ITO only include the non-Abelian Ising anyon $\sigma$ and the Majorana fermion. 
One common way to confine a topological order is to condense some of its anyonic excitations that have bosonic self-statistics and proper mutual statistics with other anyons. In an ITO, however, there is no obvious such (bosonic) anyon that can condense. 
One may also try to describe the transition in terms of gapless fermions coupled to gauge fields. 
As mentioned in the introduction, the ITO can be understood as a $Z_2$ gauge field coupled to Majorana fermions in a topological band with Chern number $C=1$. 
To confine the ITO, one needs to first change the Chern number of the Majorana fermions from $C=1$ to $C=0$. This process yields a pure deconfined $Z_2$ gauge theory, which is the more familiar $Z_2$ toric code state \cite{Kitaev2006}.
To get a topologically trivial state, one needs to further confine the pure $Z_2$ gauge theory.
In other words, one needs two separate transitions to confine the ITO.
The first transition is described by a single Majorana cone coupled to a $Z_2$ gauge field, and the second transition is the confinement transition of the pure $Z_2$ gauge theory, which can be described by an Ising order parameter coupled to a $Z_2$ gauge field \cite{Read1991}.{\footnote{We note that due to the coupling to a dynamical $Z_2$ gauge field, the first transition is in a distinct universality class compared to a single free gapless Majorana fermion, and the second tranisition is also in a distinct universality class compared to the 3D Ising transition.}}

The way to make progress, as we will discuss in the following,  is to consider dual  topological quantum field theory (TQFT) descriptions of the ITO. 
More precisely, we will find other gauge theories that are capable of describing the ITO, such that the confinement transitions of these gauge theories can be understood either by critical bosons or gapless fermions coupled to the gauge fields.

\subsection{Topological aspects and bosonic critical theories}

To apply this strategy to our case,
first recall that the ITO can be viewed as a $p+ip$ superconductor coupled to a dynamical $Z_2$ gauge field that corresponds to the fermion parity symmetry, \ie the ITO is a gauged $p+ip$ superconductor. Furthermore, there is a 16-fold-way classification of $2+1$ D gapped superconductors coupled to such a $Z_2$ gauge field, where the ITO corresponds to the state with an index $\nu=1$ \cite{Kitaev2006}. Suppose we take the superconductor with $\nu=3$ together with another superconductor with $\nu=-2$, and weakly hybridize the fermions in these two superconductors, the resulting state is the one with $\nu=1$.

This observation is useful because it is known that the state with $\nu=3$ can be described by an $SU(2)_2$ Chern-Simons theory coupled to a boson. This theory also has two nontrivial anyons: a non-Abelian anyon $\sigma'$ and a Majorana fermion. 
In addition, the state with $\nu=-2$ can be described by a $U(1)_{-4}$ Chern-Simons theory coupled to a boson, and this theory has three nontrivial Abelian anyons, with one of them a Majorana fermion. 
Therefore, we can arrive at the ITO state by taking an $SU(2)_2$ theory and a $U(1)_{-4}$ theory, and hybridizing the Majorana fermions in these two theories. 
More formally, this hybridization of the Majorana fermions can be viewed as a process of anyon condensation, where the bound state of the Majorana fermions from the $SU(2)_2$ and $U(1)_{-4}$ theories are condensed. In the language of TQFT, the resulting coupled theory is denoted as $U(2)_{2, -2}$,\footnote{In Ref.~\cite{Hsin2016}, this theory is denoted as $U(2)_{2, -2}$. In Ref.~\cite{Seiberg2016}, it is denoted as $U(2)_{2,-4}.$} and we have derived a known duality~\cite{Seiberg2016}\footnote{One can also obtain the ITO by hybridizing the Majorana fermions in a state with $\nu=n$ and those in a state with $\nu=1-n$, for other values of integral $n$. 
However, other choices of $n$ result in more complicated theories, and, as far as we know so far, the combination of $\nu=3$ and $\nu=-2$ is the only theory that agrees with the phase diagram in Fig. \ref{fig:summary}.}
\begin{equation}
\textrm{Ising TQFT} \longleftrightarrow U(2)_{2, -2}=\frac{SU(2)_2\times U(1)_{-4}}{\mathds Z_2}.
\end{equation}

The Lagrangian of the $U(2)_{2, -2}$ theory can be written as
\begin{equation}  \label{eq: U(2)-boson-top}
 \mathcal L_{\textrm{CS}} =-\frac{2}{4\pi} \textrm{Tr} ( \mathbf b d \mathbf b  - \frac{2i}{3} \mathbf b ^3) + \frac{2}{4\pi} (\textrm{Tr} \mathbf b) d (\textrm{Tr} \mathbf b).
\end{equation}
where $\mathbf b=b+\tilde b \mb{1}$ is a 2-by-2 $U(2)$ gauge field, with $b$ an $SU(2)$ gauge field and $\tilde b$ a $U(1)$ gauge field. 
\footnote{This theory can also be written as,
\begin{equation}
\mathcal L_{\textrm{CS}} = -\frac{2}{4\pi} \textrm{Tr} ( b d b - \frac{2i}{3} b^3) + \frac{4}{4\pi} \tilde b \, d \, \tilde b . \nonumber
\end{equation}
}

This $U(2)$ gauge field is coupled to dynamical bosonic matter fields $\Phi$, so that the total Lagrangian is
\begin{equation}\label{eq: U(2)-boson-total}
\mathcal L = \mathcal L_{[\Phi,\mathbf{b}]} + \mathcal L_{\textrm{CS}}+\mc{L}_{\rm Maxwell}- \frac{1}{2\pi}Bd(\textrm{Tr}\mb b)+\cdots
\end{equation}
Here $\Phi$ may have different flavors, and each flavor can be thought of as a two-component (corresponding to the color index) complex boson, $\Phi=(\phi_a,\phi_b)^T$, which are in the fundamental representation of the $U(2)$ gauge group.
The third term $\mc{L}_{\rm Maxwell}$ is the standard Maxwell Lagrangian of the gauge field, and at long distances it is less relevant compared to the topological part, Eq.~\eqref{eq: U(2)-boson-top}.

Before proceeding, we pause to comment on the global symmetries of the theory Eq. (\ref{eq: U(2)-boson-total}) in the absence of the last $\cdots$ term. As a quantum field theory in the continuum, besides the Poincare symmetry, CPT symmetry, etc., this theory also enjoys a $U(1)$ symmetry corresponding to the conservation of the gauge flux of $\tilde{b}$, as well as an $SU(N_f)$ flavor symmetry.  These symmetries may not be present in the physical system, but it is nevertheless helpful to keep track of them. The microscopic symmetries of the physical system must be embedded into these symmetries, but, a {\it priori}, the precise embedding pattern can only be determined after we have a concrete microscopic construction where this field theory emerges at long distances.  When specifying to the physical system, we will add appropriate $\cdots$ terms to Eq. (\ref{eq: U(2)-boson-total}) to break its full symmetries to the physical symmetries. For example, we can add monopole operators of $\tilde b$ to break the $U(1)$ flux conservation symmetry, and add certain quartic interactions to break this $SU(N_f)$ flavor symmetry. To keep track of the $U(1)$ symmetry, we have added the fourth term, where $B$ is the probe gauge field of this $U(1)$ symmetry.

The dynamics of the bosonic field $\Phi$ is described by the standard $\phi^4$ theory, with $N_f=1,2$ flavors,
\begin{equation}
\mathcal L_{[\Phi,\mathbf{b}]} = \sum_{I=1}^{N_f}   |(\partial_\mu - i \mathbf b_\mu) \Phi_I|^2 -m^2 \sum |\Phi_I|^2 - V(\Phi).
\end{equation}
where $V(\Phi)$ is the symmetry-consistent quartic potential term.

If the $\Phi$ fields are gapped, they are dynamically trivial and hence can be simply neglected.
The theory is then described by the $U(2)_{2,-2}$ theory, which is nothing but the ITO.
On the other hand, if the $\Phi$ fields are condensed, the $U(2)$ gauge field will be Higgsed, which destroys the ITO.
The mass of $\Phi$ is the tuning parameter for this phase transition.
In the continuum field theory, it is straightforward to understand the phases when $\Phi$ is condensed. 

When $N_f=1$, the $U(2)$ gauge field is Higgsed down to $U(1)$. Without loss of generality, let us suppose the first color component of $\Phi$ gets a nonzero vacuum expectation value, then only $b_{22}$ is an active gauge field. In the absence of the $\cdots$ term in Eq. (\ref{eq: U(2)-boson-total}), the Lagrangian describing this remaining gauge field is
\begin{align}\label{eq: phi4}
\mc{L}&=-\frac{2}{4\pi}b_{22}db_{22}+\frac{2}{4\pi}b_{22}db_{22} +\mc{L}_{\rm Maxwell} -\frac{1}{2\pi}Bdb_{22} \nonumber \\
& = \mc{L}_{\rm Maxwell}-\frac{1}{2\pi}Bdb_{22}.
\end{align}
So we end up with a $2+1$D $U(1)$ Maxwell theory, which is nothing but a Goldstone phase with the $U(1)$ flux conservation symmetry spontaneously broken \cite{Dasgupta1981}. 
In the Kitaev materials, this $U(1)$ symmetry should be explicitly broken, and the monopole operators responsible for this symmetry breaking will gap out the Goldstone mode. Physically, it may be tempting to identify this phase as the zigzag phase. However, the precise nature of this confined state depends on the quantum numbers of the monopoles, which we will discuss in the next subsection.

When $N_f=2$, the $U(2)$ gauge field will generically be completely Higgsed. 
The gauge sector is trivial, and the precise nature of the resulting confined state is determined by whether the condensate of $\Phi$ spontaneously breaks any symmetry. 
Before the condensation, the system has an flavor rotation symmetry between $\Phi_1$ and $\Phi_2$, which can maximally be $SU(2)$.
The condensation pattern of $\Phi_{1,2}$ is dependent on the form of the quartic potential $V(\Phi)$ in Eq.~\eqref{eq: phi4}.
Specifically, if $V(\Phi)$ is $SU(2)$ invariant, it should have the form $\rho \textrm{Tr} M^2 + \lambda (\textrm{Tr} M)^2$, with $M_{IJ} = \sum_a \phi_{Ia} \phi_{Ja}^\dag$. Here $I,J$ are the flavor indices, and $a$ is the color index.
If $\rho, \lambda >0$, $\Phi$ will condense in the $SU(2)$ invariant channel. In practice, the $SU(2)$ flavor symmetry is absent, but it is still possible that the condensation pattern of $\Phi$ does not break any physical symmetry, depending on the microscopic details.
So we can end up with a completely trivial state with no topological order or spontaneous symmetry breaking.

Therefore, we have reached two continuum field theories for the confinement transitions of the ITO, with $N_f=1,2$, respectively. 
In both theories, in order to determine the symmetries of the confining states, we need to understand how the physical symmetries are embedded into the emergent symmetries of Eq. (\ref{eq: U(2)-boson-total}). 
Also, we need to know whether the physical symmetries are sufficient to forbid all other possibly relevant operators with respect to these critical theories. In order to do this, a concrete microscopic construction of the critical theory is needed.
It turns out to be easier to achieve this goal with a dual fermionic description to Eq. (\ref{eq: U(2)-boson-total}), as we will discuss below.

Before leaving this subsection, we point out an interesting relation between the theory Eq. (\ref{eq: U(2)-boson-total}) and the bosonic integer quantum Hall (BIQH) states \cite{Senthil2013, Liu2013}, although this relation is not of vital relevance for the discussions in this paper. The BIQH states are often viewed as bosonic SPTs protected by a $U(1)$ symmetry, but they are in fact compatible with a $U(2)$ symmetry. These states can be labeled by their Hall conductance under the $U(1)$ gauge field corresponding to the protecting $U(1)$ symmetry, $\sigma_{xy}=2n$ with $n$ an integer (in units such that the state described in Ref. \cite{Senthil2013} has $n=1$). The response of the state with $n=-2$ to the $U(2)$ gauge field corresponding to the $U(2)$ symmetry is precisely given by Eq. (\ref{eq: U(2)-boson-top}), with the gauge fields in Eq. \eqref{eq: U(2)-boson-top} viewed as a probe gauge field \cite{Senthil2013, Liu2013}.
In other words, the ITO can be obtained by gauging two copies of the BIQH states in Ref. \cite{Senthil2013}, which is indeed similar to that the Abelian chiral spin liquid is a gauged (one-copy) BIQH state  ~\cite{Barkeshli_Chiral,YCH15_2}. In a BIQH state, the boson condensation transition is described by Eq. \eqref{eq: U(2)-boson-total} with all gauge fields taken as probe gauge fields.
Therefore, the theory Eq. \eqref{eq: U(2)-boson-total} can be understood as condensing the bosons in the gauged BIQH states, where the gauge fields are dynamical.

\subsection{Symmetry properties and dual fermionic theories} \label{subsec: fermionic dual}

The bosonic critical theory turns out to be dual to a fermionic critical theory,
\begin{align}\label{eq:fermionic}
&\mathcal L =  \sum_{I=1}^{N_f}\bar \Psi_I i(\slashed \partial - i \slashed{\mathbf a}) \Psi_I  +m \sum \bar \Psi_I \Psi_I+\mathcal L_{\textrm{top}}, \\
&\mathcal{L}_{\textrm{top}} = \frac{2-N_f/2}{4\pi} \textrm{Tr} \left[\mathbf a d \mathbf a -\frac{2i}{3}  \mathbf a^3\right]  + \left(4-N_f\right)\textrm{CS}_g \nonumber \\
&+\frac{2}{4\pi} \beta d \beta - \frac{1}{2\pi} \beta d (B-(\textrm{Tr} \mathbf a)).
\end{align}
Here $\mb a$ is a $U(2)$ gauge field, $\textrm{CS}_g$ denotes the gravitational Chern-Simons term, $\beta$ is a dynamical $U(1)$ gauge field, and $B$ is a probe gauge field of the global $U(1)$ symmetry as in the bosonic critical theory. In our convention, when the coefficient of $\gcs$ is $1$, the theory has thermal Hall conductance $\kappa_{xy}=1$ in units of $(\pi/6)(k_B^2T/\hbar)$, or, in other words, it has an edge with chiral central charge $c_-=1$.
The fermion field $\Psi$ is in the fundamental representation of the $U(2)$ gauge group, and its flavor number can be $N_f=1,2$.
This duality can be derived using the level-rank duality~\cite{Hsin2016} (see Appendix~\ref{app:level_rank}), and it was also presented in Ref.~\cite{Radicevic2016}\footnote{The duality only holds for $N_f=1,2$ ~\cite{Hsin2016}.} \footnote{The more precise form of the half-quantized Chern-Simons terms in the above Lagrangian are proper $\eta$-invariants \cite{wittenreview}. But writing the half-quantized Chern-Simons terms is more intuitive and does not alter our discussion.}. 

Here the singlet mass of Dirac fermions $m \sum \bar \Psi_I \Psi_I$ is the tuning parameter of the confinement transition.
When $m\ll -1$ (in proper units), integrating out the Dirac fermions gives a non-Abelian Chern-Simons theory,
\begin{equation}\label{eq:fermion_TQFT}
\mathcal L =\frac{2}{4\pi}  \textrm{Tr} \left[\mathbf a d \mathbf a -\frac{2i}{3}  \mathbf a^3 \right] + 4\textrm{CS}_g +\frac{2}{4\pi} \beta d \beta + \frac{1}{2\pi} \beta d (\textrm{Tr} \mathbf a).
\end{equation}
This theory indeed describes the ITO.
One might be confused about this statement, since the Chern-Simons levels here look rather distinct from those in Eq. (\ref{eq: U(2)-boson-total}).
However, it is inappropriate to directly compare the Chern-Simons levels between these two theories, because here $\mb a$ is coupled to fermions (hence it is a spin gauge field), while $\mb b$ in Eq. (\ref{eq: U(2)-boson-total}) is coupled to bosons.
After taking into account the difference in the matter fields, we can show that the topological order of Eq.~\eqref{eq:fermion_TQFT} is exactly the same as $U(2)_{2,-2}$ Chern-Simons theory, \ie the Ising TQFT (see Appendix \ref{app:level_rank}). 
We can also do a quick self-consistency check by examining the gravitational response, whose coefficient corresponds to the physical thermal Hall conductance. The non-Abelian Chern-Simons term in Eq.~\eqref{eq:fermion_TQFT} can be roughly considered as $U(2)_{-2}\times U(1)_{-2}$, and integrating out them yields a gravitational Chern-Simons term $-\frac{7}{2}\textrm{CS}_g$ ($U(2)_{-2}$ contributes $-\frac{5}{2}\textrm{CS}_g$  and $U(1)_{-2}$ contributes $-\textrm{CS}_g$).
Combined with the $4\gcs$ term in (\ref{eq:fermion_TQFT}), the total gravitational response is $\frac{1}{2}\textrm{CS}_g$, which is the identical to that of the ITO.

On the other hand, when $m\gg 1$, the ITO will be destroyed. As in the bosonic theories, the phases that ITO is confined to depend on the fermion flavor number $N_f$ and the actions of the physical symmetries in these critical theories.
More precisely, when $N_f=1$, the theory will be confined to a pure $U(1)$ Maxwell theory, in which the monopole  will proliferate and the nature of the resulting phase depends on the quantum numbers of the monopole.
When $N_f=2$, all the gauge fields will be confined without breaking any symmetry. 
This gives a trivially polarized state as long as there is no other relevant perturbation that can destroy the critical point.
To understand the final fates of the confined states, we need to have concrete microscopic constructions of these critical theories.

The above fermionic critical theories motivate a parton construction for the ITO and its confinement transitions.
With such an explicit construction, we are able to directly work out the symmetry properties of the field theories. 
In particular, using our parton constructions, we will show that, 
\begin{itemize}
\item[]
i) The confined phase in the theory with $N_f=1$ can indeed be the zigzag magnetic order.
\item[]
ii) The confined phase in the theory with $N_f=2$ can indeed be a trivial state with all symmetries preserved.
\item[]
iii) The symmetries of the representative Kitaev materials (listed in Table \ref{tab:symmetry}) are sufficient to forbid the most obvious relevant operators in both critical theories (with $N_f=1$ and $N_f=2$, respectively). 
\end{itemize}

To make the symmetries of Kitaev materials manifest, we consider a rotated spin basis,
\beq \label{eq: new spins}
\begin{split}
\widetilde S^x &=\frac{S^x+S^y+S^z}{\sqrt 3},  \\
\widetilde S^y &=\frac{S^x+S^y-2S^z}{\sqrt 6}, \\
\widetilde S^z &=\frac{S^x-S^y}{\sqrt 2}.
\end{split}
\eeq
Here $\widetilde S^{x,y,z}$ are chosen to be parallel to the $c^*$, $a$ and $b$ axes, so that they have simple symmetry transformation rules. For example, under $\sigma^*$, instead of transforming as in Eq. \eqref{eq: action of sigma*}, they transform as
\beq
\begin{split}
    &\tilde S^x_{\vec r}\rightarrow -\tilde S^x_{\sigma\vec r}\\
    &\tilde S^y_{\vec r}\rightarrow -\tilde S^y_{\sigma\vec r}\\
    &\tilde S^z_{\vec r}\rightarrow \tilde S^z_{\sigma\vec r}
\end{split}
\eeq

The parton construction is~\cite{Wen_nonAbelianFQH}:
\begin{equation}\label{eq:QCD_parton}
\widetilde S^+ = \phi^\dag f_a^\dag f_b^\dag, \quad \quad \widetilde S^z = \frac{n_\phi + n_{f_a} + n_{f_b}}{3} - \frac 1 2,
\end{equation}
with a constraint $n_\phi = n_{f_a} = n_{f_b}$.
This parton construction has a $U(2)$ gauge invariance: $\Psi = (f_a, f_b)^T$ is in the $U(2)$ fundamental representation, and it is interacting with a $U(2)$ gauge field $\mb a$;  $\phi$ carries charge under the diagonal part of the $U(2)$ gauge field and a global $U(1)$ charge (of the $\tilde S_z$ rotation). 

To get the ITO, we can put the bosonic parton $\phi$ into a Laughlin state at $\nu=-1/2$, and put the fermionic partons $f_i$ into a topological band with Chern number $C=2$. 
This gives exactly the Chern-Simons theory in Eq.~\eqref{eq:fermion_TQFT}: the fermionic parton contributes a $U(2)_{-2}$ Chern-Simons term for $\mb a$, while the bosonic parton is described by a $U(1)_{-2}$ Chern-Simons term of the $U(1)$ gauge field $\beta$.

The confinement transition of ITO can be triggered by changing the Chern number of fermionic partons.
Specifically, for a transition from $C=2$ to $C=1$, we get a critical theory with $N_f=1$, while for a transition from $C=2$ to $C=0$, we get a critical theory with $N_f=2$.
In Appendix~\ref{app:U(2)-parton_fermionic}, we provide the concrete mean-field ansatzs for these two Chern-number-changing transitions.
We only consider a Zeeman field on the $ac^*$ plane, which is the direction of Zeeman field reported in Ref.~\cite{Rucl3_kappaxy}. 
In this case, the symmetries of the system include translation $T_{1,2}$, inversion $C_2$, as well as the combination of time-reversal and pseudo-mirror $\mathcal T \sigma^*$ (see Table~\ref{tab:symmetry}).
We also work out how those symmetries are implemented in the critical theories using the mean-field ansatzs.

In the theory with $N_f=1$, besides the singlet mass (tuning parameter of the transition), the most relevant operators are the monopole operator $\mathcal M$, conserved current, $d\left(\textrm{Tr}\mb a\right)$, and $\bar \Psi \gamma^\mu \Psi$ \footnote{The scaling dimension of $\bar\Psi\gamma^\mu\Psi$ in the presence of a Chern-Simons term, as is the case here (for both $N_f=1$ and $N_f=2$), is $2$, since it has identical symmetry quantum numbers as $d(\Tr\mb{a})$, which has scaling dimension $2$. In the absence of a Chern-Simons term, its scaling dimension is $3$ \cite{hermele2005}.}. Their quantum numbers are shown in Table~\ref{tab:symmetry_Nf1}, and all of them are disallowed by symmetries.
The minimally allowed monopole operator is a two-fold monopole, which may or may not be relevant in the infrared. 
If it is irrelevant, we may have a stable critical point with an emergent $U(2)$ gauge field.
Furthermore, the monopole has exactly the same quantum number as the zigzag magnetic order, assuming the magnetic moments are ordered on the $ac^*$ plane in the zigzag phase, as suggested by Refs. \cite{Cao2016, Chun2015}. 
Therefore, the proliferation of monopoles in the theory with $N_f=1$ results in precisely the zigzag magnetic order.

\begin{table}
\setlength{\tabcolsep}{0.3cm}
\renewcommand{\arraystretch}{1.4}
    \centering
    \begin{tabular}{c|ccccc} 
    \hline \hline
    & $T_{1}$ & $T_2$ & $C_2$ & $\mathcal T \sigma^*$ \\ \hline
  $\mathcal M$  & $-1$ & $-1$ & $-1$ & $-\mathcal M^\dag$ \\       
$\bar \Psi \gamma^0 \Psi$ & $1$ & $1$ & $1$ & $-1$ \\ 
$\bar \Psi \gamma^1 \Psi$ & $1$ & $1$ & $-1$ & $1$ \\ 
$\bar \Psi \gamma^2 \Psi$ & $1$ & $1$ & $-1$ & $-1$ \\ 
\hline\hline
    \end{tabular}
    \caption{Symmetries of operators in the $N_f=1$ critical theory. $d\left(\textrm{Tr}\mb a\right)$ happens to have the same quantum number as $\bar \Psi \gamma^\mu \Psi$.}
    \label{tab:symmetry_Nf1}
\end{table}

\begin{figure}
    \centering
    \includegraphics[width=0.5\linewidth]{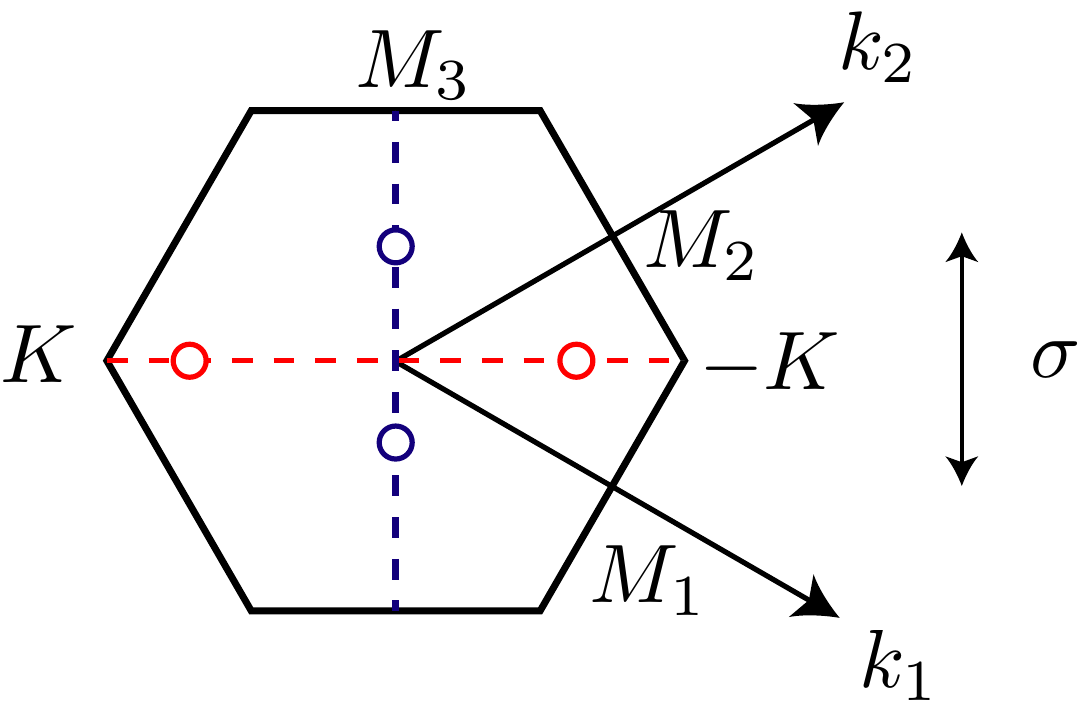}
    \caption{Brillouin zone of the honeycomb lattice. 
    The red  dashed line represents the high-symmetry line which connects the $K$ and $K'=-K$ points; 
    The blue dashed line connects the $M_3$ and $M_3$ point.}
    \label{fig:QCD_BZ}
\end{figure}

We now turn to the critical theory with $N_f=2$. 
Besides the $U(1)$ flux conservation, the critical theory also has an $SU(2)$ flavor rotation symmetry.
The most relevant operators are the monopole operator $\mathcal M_{1,2,3}$, $SU(2)$ adjoint mass $\bar \Psi \tau^\alpha \Psi$($\tau$ acts on the flavor index),  and conserved currents, $d(\textrm{Tr}\mb a)$, $\bar \Psi \gamma^\mu \tau^{\alpha} \Psi$ and $\bar \Psi \gamma^\mu \Psi$.
Here the monopoles are in the adjoint representation of the $SU(2)$ flavor symmetry, and it has three components. Again, we want to work out the quantum numbers of these operators to see if they are forbidden by symmetries.
There turn out to be three different cases, depending on the locations of Dirac cones.
Constrained by symmetries, the two Dirac points have to stay at the high-symmetry points/lines (see  Fig.~\ref{fig:QCD_BZ}):
\begin{enumerate}
    \item The two Dirac cones are at the $M_{1,2}$ points, $(k_1, k_2)=(\pi, 0), (0,\pi)$.
    \item The two Dirac cones are at $(k_1, k_2)=(k, k), (-k, -k)$ points ($k$ is an arbitrary number), which are on the high-symmetry line $K-K'$.
    \item The two Dirac cones are at $(k_1, k_2)=(k, -k), (-k, k)$ points ($k$ is an arbitrary number), which are on the high-symmetry line $M_3-M_3$.
\end{enumerate}
In Appendix~\ref{app:U(2)-parton_fermionic}, we provide the mean-field ansatzs for all the three possibilities, and the quantum numbers of operators are summarized in Table~\ref{tab:symmetry_Nf2_M12}-\ref{tab:symmetry_Nf2_M3}.

In case (1) (the nodes are located at $M_{1,2}$ points), there is one symmetry allowed operator, $\bar \Psi \gamma^0 \tau^z \Psi=\Psi^\dag_1 \Psi_1-\Psi^\dag_2 \Psi_2$.
This operator will destabilize the quantum critical point: it will dope the Dirac cones at the $M_1, M_2$ points and generate particle and hole pockets. 
These two Fermi pockets are interacting with a $U(2)$ gauge field, which may or may not be stable.
In cases (2) and (3), again there is one symmetry allowed relevant operator in each case: $\bar \Psi \gamma^{1} \tau^z \Psi$ and $\bar \Psi \gamma^{2} \tau^z \Psi$, respectively.
However, different from the first situation, these operators will not destroy the quantum critical points.
Instead, they will just move the Dirac points along the high symmetry lines (along either the $K$-$K'$ or the $M_3$-$M_3$ line).
Therefore, the quantum critical point between the ITO and polarized state may be stable  if the two Dirac nodes are staying at the high symmetric lines.

\section{Experimental signatures} \label{sec: exp}

So far we have theoretically explored novel  QCD$_3$ quantum phase transitions in the Kitaev materials. In this section we discuss their experimental signatures.

It is the most important to first establish experimentally that such phase transitions are indeed continuous. We remark that even if such phase transitions are continuous, naively one would not expect them to be described by $2+1$ dimensional conformal field theories (CFTs),{\footnote{A classic example of a continuous quantum phase transition which is not described by a CFT is the superfluid-insulator transition in a Bose-Hubbard model without fixing the boson density \cite{Sachdev_book2011}.}} as we propose. This is because for a transition to be described by a CFT, there often need to be many symmetries to prohibit relevant perturbations that would destabilize the CFT. However, the field-induced transitions considered here enjoy very few symmetries. Therefore, verifying that these transitions are described by CFTs already provide a nontrivial check of our theory.

If the transitions can be confirmed to be continuous and they are described by 2+1 dimensional CFTs as we propose, the qualitative behaviors of many physical quantities are readily determined and can thus be used to verify the transitions are indeed described by CFTs. For example, for a 2+1 dimensional CFT, both the specific heat and the thermal conductivity tensor behave as $T^2$ in the low-temperature limit. More generally, define
\beq
\kappa_{\alpha\beta}=\frac{1}{T^2}\la s_\alpha(k=0)s_\beta(-k=0)\ra
\eeq
with $\alpha, \beta=0,1,2$. Here $s_0$ is the energy density and $s_i$ with $i=1,2$ is the energy current. $k=(\omega,\vec k)$ collectively denotes the frequency and wave vector. $\kappa_{00}$ is the specific heat, and $\kappa_{ij}$ is the thermal conductivity tensor \cite{McLennan1959, McLennan1960, William1962}. For a 2+1 dimensional CFT, $\kappa_{\alpha\beta}$ obeys the scaling form
\beq
\kappa_{\alpha\beta}(T, B)=T^2\tilde\kappa_{\alpha\beta}(T/|B-B_c|^{\nu})
\eeq
where $\tilde\kappa_{\alpha\beta}$ is a universal function, $B_c$ is the critical field strength of these field-induced phase transitions, and $\nu$ is the critical exponent governing the divergence of the correlation length upon approaching the critical point, \ie $\xi\sim|B-B_c|^{-\nu}$ with $\xi$ the correlation length. Using the above scaling relation, by experimentally measuring the specific heat and thermal conductivity tensor in the vicinity of the quantum critical points, one can verify that these transitions are described by CFTs and obtain the correlation length exponent, $\nu$.

Using the symmetry properties of various operators in Tables \ref{tab:symmetry_Nf1}, \ref{tab:symmetry_Nf2_M12} and \ref{tab:symmetry_Nf2_K}, we can predict at which momenta of the BZ strong signals of critical modes will appear in neutron scattering experiments at these critial points. Specifically, we will consider operators (in these tables) that will potentially show a divergent peak in the neutron scattering signals as a function of momentum, and we will determine the locations of these peaks in the BZ. Below we summarize the relevant results, and the details can be found in Appendix \ref{app: neutron scattering}.

\begin{figure}[h]
    \centering
    \includegraphics[width=0.4\linewidth]{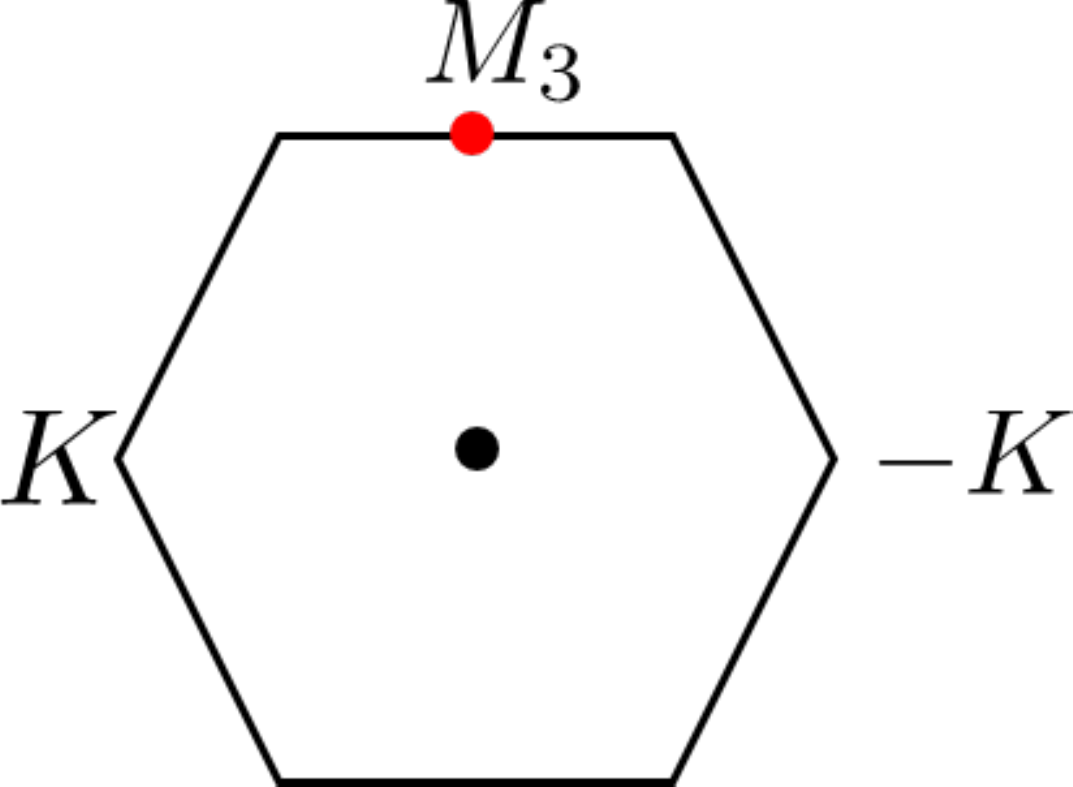}
    \caption{At the transition between the zigzag phase and the ITO phase, strong signals of neutron scattering are expected to appear at the $M_3$ point (the red point) in the BZ. There can also be a strong signal at the $\Gamma$ point (the black point). The signal at the $M_3$ point is due to the monopole operator $\mathcal{M}$, and the one at the $\Gamma$ point is due to the fermion bilinear operator $\bar\Psi\Psi$ (see Appendix \ref{app: neutron scattering} for more details).}
    \label{fig:zigzag-ITO}
\end{figure}

For the transition between the zigzag phase and the ITO phase, a peak in the spin structure factor $\la\widetilde S_{i}(\omega=0, \vec k)\widetilde S_i(\omega=0, -\vec k)\ra$ may appear at the $M_3$ point of the BZ, {\footnote{Notice, in order to have a peak in the spin structure factor, we have assumed certain operators at the critical points have a scaling dimension smaller than $3/2$. If their scaling dimension is larger than $3/2$, at these momenta the spin structure factor should show a dip rather than a peak. See Appendix \ref{app: neutron scattering} for more details.}} where $i=x, y$ (see Fig. \ref{fig:zigzag-ITO}). The critical exponent characterizing the divergence of this spin structure factor when $\vec k$ approaches $M_3$ is independent of $i=x,y$, due to an emergent $U(1)$ spin rotational symmetry around $\widetilde S_z$ in our critical theory.{\footnote{There can in principle also be strong signals at the $\Gamma$ point, as noted in Fig. \ref{fig:zigzag-ITO}. Although in real experiments such signals may be hard to extract due to the background signals from neutrons that are not scattered, but they can in principle be detected in numerical studies.}}

\begin{figure}[h]
    \centering
    \includegraphics[width=\linewidth]{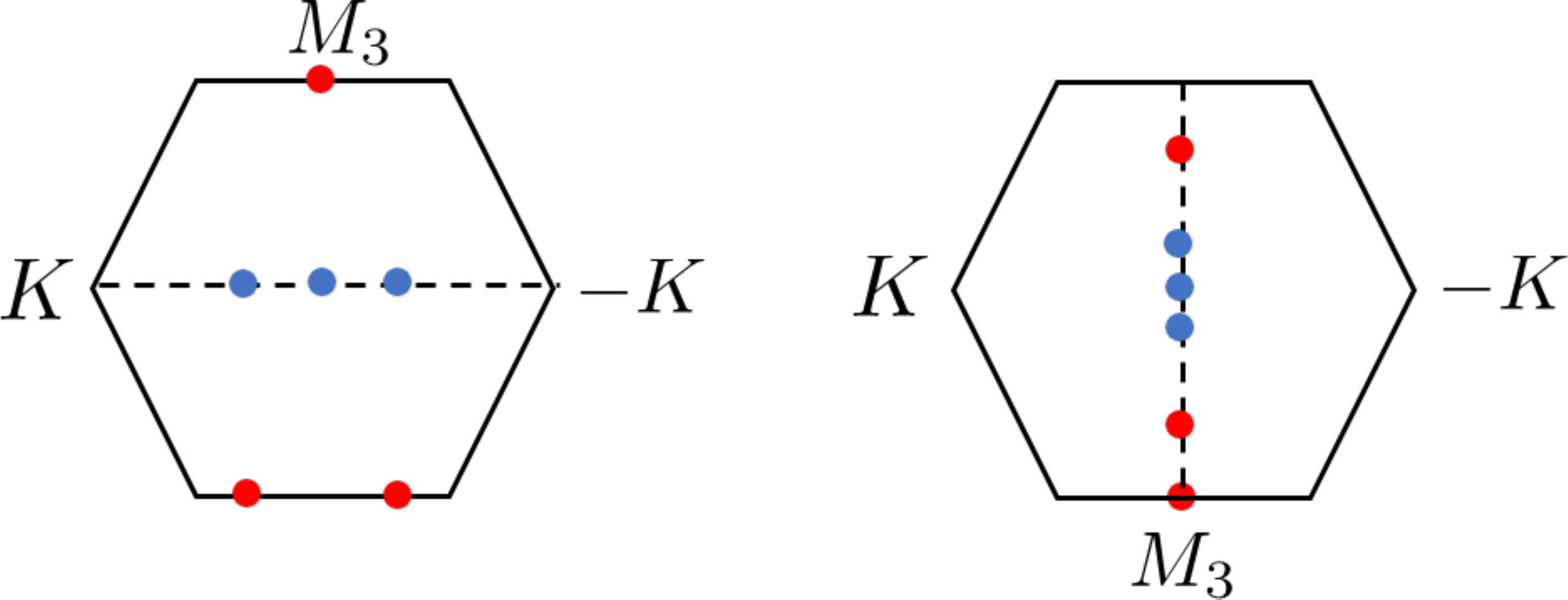}
    \caption{At the transition between the ITO phase and the trivial phase, strong signals of neutron scattering are expected to appear at the colored points in the BZ. The points with the same color have the same critical exponent of the neutron scattering signal $\la\widetilde S_{i}(\omega=0, \vec k)\widetilde S_i(\omega=0, -\vec k)\ra$. Left: for the critical theory described by case-2 in Sec. \ref{subsec: fermionic dual}, the three blue points are located at $\Gamma$, $(2k, 2k)$ and $(-2k, -2k)$, respectively. The three red points are at $M_3$, $(2k+\pi, 2k+\pi)$ and $(-2k+\pi, -2k+\pi)$, respectively. Right: for the critical theory described by case-3 in Sec. \ref{subsec: fermionic dual}, the three blue points are located at $\Gamma$, $(2k, -2k)$ and $(-2k, 2k)$, respectively. The three red points are at $M_3$, $(2k+\pi, -2k+\pi)$ and $(-2k+\pi, 2k+\pi)$, respectively. There can also be another strong signal at the $\Gamma$ point (not shown in the figure above) due to the operator $\bar\Psi\Psi$ (see Appendix \ref{app: neutron scattering} for more details).}
    \label{fig:ITO-trivial}
\end{figure}

For the transition between the ITO phase and the trivial phase, there may also be peaks in the spin structure factor $\la\widetilde S_i(\omega=0, \vec k)\widetilde S_i(\omega=0, -\vec k)\ra$, and their locations are shown as colored points in Fig. \ref{fig:ITO-trivial}. At each point, the dependence on the spin polarizations of the critical exponent characterizing how fast the corresponding peak diverges is detailed in Appendix \ref{app: neutron scattering}. Furthermore, due to an emergent $SO(3)$ flavor symmetry, points with the same color have the same critical exponent.  

These spin structure factors can be measured by neutron scattering experiments, and they provide highly-nontrivial checks of our critical theories. Notice to examine the emergent $U(1)$ symmetries at these transitions, spin-polarized neutron scattering experiments are needed. Otherwise, spin-unpolarized ones are sufficient to check the above predictions, including the emergent $SO(3)$ symmetry at the transition between ITO and the polarized state.

Another prediction of our theory is that, if the Zeeman field is titled away from the $ac^*$ plane (corresponding to breaking $\mathcal T \sigma^*$ symmetry), our QCD$_3$ quantum critical points will be unstable to either a first order phase transition or a new intermediate phase.

\section{Summary and discussion} \label{sec: discussion}

Motivated by the recent theoretical and experimental progress in the research on Kitaev materials, we study novel field-induced quantum phase transitions in these materials.
In particular, based on general symmetry grounds, we have discussed the transitions from the Ising topologically ordered (ITO) state to the zigzag order and to the trivial polarized state. We find that these transitions are rather exotic, and they can be described by QCD$_3$-Chern-Simons theories. More precisely, the transition between the ITO state and the zigzag order (the trivial polarized state) can be described by a dynamical $U(2)$ gauge field coupled to $N_f=1$ ($N_f=2$) critical fermions. We have checked that the symmetries of some representative Kitaev materials (listed in Table \ref{tab:symmetry}) are sufficient to forbid the most obvious relevant operators (other than the transition tuning operator) of these putative critical theories. Therefore, these transitions can potentially be generic direct continuous quantum phase transitions. We note that our method can also be adopted to study the transitions between the ITO and magnetic orders other than the zigzag type.

We also notice that these critical theories are dual to $N_f=1$ ($N_f=2$) species of critical bosons coupled to a dynamical $U(2)$ gauge field. 
There is an interesting relation between these critical theories with bosonic integer quantum Hall (BIQH) states. The quantum phase transitions from the BIQH states to a superfluid and to a trivial insulator have been widely studied in recent years~\cite{Tarun2013_QHtransition,LuLee,WangDCPdual}, and it may be worth relating these transitions to the transitions from the ITO state to other states.

We emphasize that our discussion on the QCD$_3$-Chern-Simons quantum criticalities is very general, and is independent of the microscopic details (\eg form of the spin-spin interactions) of the Kitaev materials. Whether these critical points are realized in a particular Kitaev material again should be determined experimentally. Some experimental signatures of these phase transitions are discussed in Sec. \ref{sec: exp}.

As for future directions, it is worth studying these quantum phase transitions in more depth. On the experimental front, it is helpful to examine the phase diagrams of the Kitaev materials more closely, and identify different phases and study the phase transitions. Numerically, it is important to study the phase diagram of more realistic lattice models. 
In the purely theoretical direction, it will be of interest to study the low-energy dynamics of these critical theories to determine whether these transitions can indeed be continuous, and what the critical exponents are. These studies will provide further insights on the experimental studies of the Kitaev materials. Also, given the similarities among the critical theories between different pairs of phases, it may be interesting to look for a theory of a multi-critical point that becomes these phases and critical theories upon adding perturbations. This will potentially lead to unified understanding of the rich structures of the quantum magnetism in these systems.

Finally, we remark that dualities and emergent non-Abelian gauge theories similar to ours may be useful tools to understand other types of exotic quantum phases and phase transitions in condensed matter systems, and we expect more applications of related ideas will arrive in the future and prove helpful.

\section{Acknowledgement}

We are grateful for inspiring discussions with Lukas Janssen, Itamar Kimchi, Sung-Sik Lee, Max Metlitski, T. Senthil, Ashvin Vishwananth, Chong Wang and Linda Ye.
L. Z. was supported by NSF grant DMR-1608505. L. Z. thanks Perimeter Institute for Theoretical Physics for hospitality, where part of this work was done. This research was supported in part by Perimeter Institute for Theoretical Physics. Research at Perimeter Institute is supported by the Government of Canada through the Department of Innovation, Science and Economic Development Canada and by the Province of Ontario through the Ministry of Research, Innovation and Science.

{\it Note added:} In a previous version of this paper on arXiv, we discussed both the exotic QCD$_3$-Chern-Simons quantum phase transitions and a gapless phase in the Kitaev model supplemented with a magnetic field. Only the former is in the current paper, and the latter will be discussed in another separate paper.

\bibliography{NFS-QCD.bib,FCI.bib,CFT.bib}

\clearpage
\onecolumngrid
\appendix

\section{Derivation of duality of critical theories}
\label{app:level_rank}

We can use the level-rank duality in Ref.  \cite{Hsin2016} to show that the bosonic critical theory Eq.~\eqref{eq:  U(2)-boson-total} is dual to the fermionic critical theory Eq.~\eqref{eq:fermionic}.
We begin with a level-rank duality, namely, that the $U(2)_{2}$ theory with $N_f$ fundamental bosons is dual to the $SU(2)_{-2+N_f/2}$ theory with $N_f$ fundamental fermions.
The duality only holds for $N_f=1,2$.
The bosonic theory is, 
\begin{equation}
\mathcal L = \sum_{I=1}^{N_f} |(\partial_\mu - i \mathbf b_\mu) \Phi_I|^2 -m^2 \sum |\Phi_I|^2 - V(|\Phi|) - \frac{2}{4\pi} \textrm{Tr} ( \mathbf b d \mathbf b  - \frac{2i}{3} \mathbf b ^3) -\frac{1}{2\pi} B' d (\textrm{Tr} \mathbf b).
\end{equation}
And the fermionic dual is, 
\begin{equation}
\mathcal L = \sum_{I=1}^{N_f}  \bar \Psi_I (i\slashed \partial + \slashed a + \frac{\slashed B'}{2} \mathbf 1_2 + m) \Psi_I  + \frac{2-N_f/2}{4\pi} \textrm{Tr} \left[(a+\frac{B'}{2} \mathbf 1_2) d (a+\frac{B'}{2} \mathbf 1_2) -\frac{2i}{3}  (a+\frac{B'}{2} \mathbf 1_2)^3 \right]+ (4-N_f)\textrm{CS}_g,.
\end{equation}
Here $\mb b$ is a $U(2)$ gauge field, $a$ is an $SU(2)$ gauge field, and $B'$ is a $U(1)$ probe field.
$V(\Phi)$ is the $SU(N_f)$-invariant quartic term.

Next we add a TQFT $U(1)_{-2}$ to both theories, yielding two new theories that are dual to each other.
The bosonic theory changes to
\begin{equation}
\mathcal L = \sum_{I=1}^{N_f} |(\partial_\mu - i \mathbf b_\mu) \Phi_I|^2 -m \sum |\Phi_I|^2 - V(|\Phi|) - \frac{2}{4\pi} \textrm{Tr} ( \mathbf b d \mathbf b  - \frac{2i}{3} \mathbf b ^3) -\frac{1}{2\pi} B' d (\textrm{Tr} \mathbf b) +\frac{2}{4\pi} \beta d \beta - \frac{1}{2\pi} \beta d (B-B'),
\end{equation}
And the fermionic theory changes to,
\begin{align}
\mathcal L = &\sum_{I=1}^{N_f}  \bar \Psi_I (i\slashed \partial + \slashed a + \frac{\slashed B'}{2} \mathbf 1_2 + m) \Psi_I  + \frac{2-N_f/2}{4\pi} \textrm{Tr} \left[(a+\frac{B'}{2} \mathbf 1_2) d (a+\frac{B'}{2} \mathbf 1_2) -\frac{2i}{3}  (a+\frac{B'}{2} \mathbf 1_2)^3 \right]+ (4-N_f)\textrm{CS}_g \nonumber \\ &+ \frac{2}{4\pi} \beta d \beta - \frac{1}{2\pi} \beta d (B-B') .
\end{align}

At last, we gauge the $U(1)$ probe field $B'\rightarrow \alpha$. 
In the bosonic theory, we can simply integrate out $\alpha$, yielding $\beta = \textrm{Tr} \mathbf b$, and the theory exactly reduces to the bosonic critical theory  Eq.~\eqref{eq: U(2)-boson-total} we introduced in the main text:
\begin{equation}
\mathcal L = \sum_{I=1}^{N_f} |(\partial_\mu - i \mathbf b_\mu) \Phi_I|^2 -m \sum |\Phi_I|^2 - V(|\Phi|) - \frac{2}{4\pi} \textrm{Tr} ( \mathbf b d \mathbf b  - \frac{2i}{3} \mathbf b ^3)  +\frac{2}{4\pi} (\textrm{Tr} \mathbf b)  d (\textrm{Tr} \mathbf b) - \frac{1}{2\pi} B d (\textrm{Tr} \mathbf b),
\end{equation}
In the fermionic theory, gauging $B'$ will promote $a+ \frac{B'}{2} \mathbf 1_2$ to a $U(2)$ gauge field $\mathbf a$, 
\begin{equation}\label{eq:app_fermionic}
\mathcal L = \sum_{I=1}^{N_f}  \bar \Psi_I (i\slashed \partial + \slashed{\mb a} + m) \Psi_I  + \frac{2-N_f/2}{4\pi} \textrm{Tr} \left[ \mb a d \mb a -\frac{2i}{3}  \mb{a}^3 \right]+ (4-N_f)\textrm{CS}_g + \frac{2}{4\pi} \beta d \beta - \frac{1}{2\pi} \beta d (B-(\textrm{Tr} \mathbf a)).
\end{equation}
This theory is exactly the fermionic critical theory Eq.~\eqref{eq:fermionic} we introduced in the main text.

In the rest of this appendix we will derive the topological nature of the resulting phases for different signs of $m$ and for $N_f=1, 2$. For this purpose, it is sufficient to switch off the probe gauge field $B$, \ie we will set $B=0$. The method presented below can be straightforwardly adopted to determine the topological nature of similar theories.

\subsection*{$m\ll -1$ with $N_f=1, 2$: ITO}

Let us start with $m\ll -1$. In this case, integrating out the fermions results in the following effective Lagrangian (for both $N_f=1$ and $N_f=2$):
\beq
\begin{split}
\mc{L}
&=\frac{2}{4\pi}\Tr
\left[
\mathbf{a}d\mathbf{a}-\frac{2i}{3}\mathbf{a}^3\right]+4\gcs+\frac{2}{4\pi}\beta d\beta+\frac{1}{2\pi}\beta d(\Tr\mathbf{a})\\
&=\frac{2}{4\pi}\Tr\left[ada-\frac{2i}{3}a^3\right]+\frac{4}{4\pi}\tilde ad\tilde a+\frac{2}{4\pi}\beta d\beta+\frac{2}{2\pi}\beta d\tilde a+4\gcs
\end{split}
\eeq
where $\mathbf{a}=a+\tilde a\mb{1}$, with $a$ an $SU(2)$ gauge field and $\tilde a$ a $U(1)$ gauge field.

Now we would like to understand why this Chern-Simons-matter theory describes the ITO state, \ie a topological order with anyon contents $\{1, \sigma, \epsilon\}$, where $\sigma$ is a non-Abelian anyon with topological spin $\theta_\sigma=e^{i\frac{\pi}{8}}$ and $\epsilon$ is a Majorana fermion. To this end, let us first understand different sectors of this theory. Denote the Lagrangian of the first sector by $\mc{L}_1$:
\beq
\mc{L}_1=\frac{2}{4\pi}\Tr\left[ada-\frac{2i}{3}a^3\right]
\eeq
and the Lagrangian of the second sector by $\mc{L}_2$:
\beq
\mc{L}_2=\frac{4}{4\pi}\tilde ad\tilde a+\frac{2}{4\pi}\beta d\beta+\frac{2}{2\pi}\beta d\tilde a
\eeq

If $\mc{L}_1$ described a Chern-Simons field coupled to bosonic matter fields, it was precisely $SU(2)_{-2}$, \ie it described a topological order with anyon content $\{1, \sigma_{-3}, \epsilon\}$, where $\sigma_{-3}$ is a non-Abelian anyon with topological spin $\theta_{\sigma_{-3}}=e^{-i\frac{3\pi}{8}}$, and $\epsilon$ is a Majorana fermion. This is also Kitaev's $\nu=-3$ state in the 16-fold way \cite{Kitaev2006}. In terms of the Chern-Simons-matter field theory, the $\sigma_{-3}$ excitation is obtained by exciting a matter field in the spinor representation of the $SU(2)$ gauge group, then the Chern-Simons term will associate some $SU(2)$ flux to this excitation and convert it into the non-Abelian anyon $\sigma_{-3}$. Importantly, here our Chern-Simons gauge field $a$ is coupled to a fermionic matter in the fundamental representation, and the fermionic nature of the matter field will change the topological spin of this excitation from $e^{-i\frac{3\pi}{8}}$ to $e^{-i\frac{3\pi}{8}}\times(-1)=e^{i\frac{5\pi}{8}}$. Let us suggestively denote this excitation as $\sigma_5$. The $\epsilon$ excitation is obtained by exciting a matter field in the integer-spin representation of the $SU(2)$ gauge field. Since matter fields in such representations are bosonic, the topological spin of this excitation will not be modified. One can further check the fusion and braiding, and verify that $\mc{L}_1$ coupled to fermionic matter in the fundamental representation is actually the topological order with $\nu=5$ in Kitaev's 16-fold way (up to the chiral central charge on the edge), with anyon content $\{1, \sigma_5, \epsilon\}$ \cite{Kitaev2006}.

Next we examine the property of $\mc{L}_2$. Using the standard K-matrix formalism \cite{Wen2004Book}, the topological nature of this theory can be determined by first rewriting $\mc{L}_2$ as
\beq
\mc{L}_2=\frac{K_{IJ}}{4\pi}a_Ida_J
\eeq
where $a_I=(\tilde a, \beta)^T$ and
\beq
K_{IJ}=
\left(
\begin{array}{cc}
4 & 2\\
2 & 2
\end{array}
\right)
\eeq
To read off the topological properties of this state, we need to invert the matrix $K$ and get
\beq
K^{-1}=
\left(
\begin{array}{cc}
\frac{1}{2} & -\frac{1}{2} \\
-\frac{1}{2} & 1
\end{array}
\right)
\eeq

The excitations of this theory can be labeled by an excitation vector $l$, and the elementary ones are $l_1=(1, 0)^T$ and $l_2=(0, 1)^T$. If the Chern-Simons gauge fields in $\mc{L}_2$ are coupled to bosonic matter fields, the excitation labeled by $l_1$ has topological spin $\theta_1=e^{-i\frac{\pi}{2}}$, and the excitation labeled by $l_2$ has topological spin $\theta_2=-1$. These two excitations, $l_1$ and $l_2$, have mutual braiding $\theta_{l_1l_2}=-1$.

Notice that $l_1$ carries charge-1 under both $\tilde a$ and $a$, so the excitation associated with $l_1$ is actually bound with $\sigma_5$ in the sector of $\mc{L}_1$. This composite excitation has topological spin $e^{i\frac{5\pi}{8}}\times e^{-i\frac{\pi}{2}}=e^{i\frac{\pi}{8}}$, and it will be identified as the $\sigma$ excitation in the ITO. At this point, there seem to be three nontrivial topological excitations: $\sigma$, $\epsilon$ and excitation $l_2$ in the sector $\mc{L}_2$. As argued before, $l_2$ is a fermion. In fact, $l_2$ should be identified with $\epsilon$. To see this, consider the bosonic bound state $\epsilon\cdot l_2$. It is straightforward to check that this bound state has no nontrivial braiding with all other excitations. Therefore, this excitation must be local. In other words, $\epsilon$ and $l_2$ are in fact in the same topological sector. 

In summary, the final anyon content is $\{1, \sigma, \epsilon\}$, which is precisely the same anyon content as the ITO state. Also, the fusion and braiding properties of these excitations also match with ITO. Furthermore, in Sec. \ref{sec: QCD3} we have verified that the chiral central charge of the edge states of this theory matches with that of the ITO. Therefore, we conclude that the theory described by $m\ll -1$ is precisely the ITO state, for both $N_f=1$ and $N_f=2$.

\subsection*{$m\gg 1$ with $N_f=1$: a short-range entangled state}

Next, let us move to the case with $m\gg 1$ and $N_f=1$. In this case, integrating out the fermions leads to the following effective Lagrangian:
\beq
\mc{L}=\frac{1}{4\pi}\Tr\left[ada-\frac{2i}{3}a^3\right]+\frac{2}{4\pi}\tilde ad\tilde a+\frac{2}{4\pi}\beta d\beta+\frac{2}{2\pi}\beta d\tilde a+2\gcs
\eeq

Again, let us look at the two sectors separately:
\beq \label{eq: positive m, Nf=1}
\begin{split}
\mc{L}_1&=\frac{1}{4\pi}\Tr\left[ada-\frac{2i}{3}a^3\right]\\
\mc{L}_2&=\frac{2}{4\pi}\tilde ad\tilde a+\frac{2}{4\pi}\beta d\beta+\frac{2}{2\pi}\beta d\tilde a
\end{split}
\eeq

In the sector described by $\mc{L}_1$, if the Chern-Simons gauge field is coupled to a bosonic matter, it is precisely the $SU(2)_{-1}$ theory, which has a topological order with only one nontrivial excitation, an anti-semion $\bar{s}$ with topological spin $\theta_{\bar{s}}=e^{-i\frac{\pi}{2}}$. Again, this excitation comes from exciting a matter field in the spinor representation of the $SU(2)$ gauge field. Because here the spinor representations are all fermionic, this anti-semion will be converted into a semion $s$ with topological spin $\theta_s=e^{i\frac{\pi}{2}}$.

In the sector described by $\mc{L}_2$, let us first define $a_\pm=\frac{1}{2}(\tilde a\pm\beta)$. In terms of $a_\pm$, $\mc{L}_2$ can be written as
\beq
\mc{L}_2=\frac{8}{4\pi}a_+da_+
\eeq
Notice the absence of a Chern-Simons term for $a_-$ here, which means this gauge field should be confined due to monopole proliferation. That is to say, the deconfined excitation in the sector described by $\mc{L}_2$ should have zero charge under $a_-$. It is straightforward to verify that these excitations all have even charges under $a_+$. So the elementary nontrivial excitation in this sector is given by having charge-2 under $a_+$, and this excitation has topological spin $\theta=e^{-i\frac{\pi\times 2\times 2}{8}}=e^{-i\frac{\pi}{2}}$. It is also easy to see that any excitation with charge-$2$ under $a_+$ also carries a spinor representation of the $SU(2)$ gauge field $a$ in the sector of $\mc{L}_1$, so this excitation is always bound with the $s$ excitation from $\mc{L}_1$, and the resulting composite excitation is a boson. One can verify there is no other nontrivial excitation in this theory, which means the topological order is actually trivial. Furthermore, integrating out the gauge field in $\mc{L}_1$ of Eq. (\ref{eq: positive m, Nf=1}) generates $-\gcs$, and integrating out the gauge field in $\mc{L}_2$ of Eq. (\ref{eq: positive m, Nf=1}) also generates $-\gcs$. Adding them together cancels the background term $2\gcs$, so the resulting state has vanishing chiral central charge on the edge.

In summary, the case with $m\gg 1$ and $N_f=1$ is a short-range entangled state, \ie it has no nontrivial topological excitation or nontrivial edge mode.

\subsection*{$m\gg 1$ and $N_f=2$: a short-range entangled state}

Finally, let us turn to the case with $m\gg 1$ and $N_f=2$. In this case, integrating out the fermions gives rise to the following effective Lagrangian:
\beq \label{eq: positive m, Nf=2}
\mc{L}=\frac{2}{4\pi}\beta d\beta+\frac{2}{2\pi}\beta d\tilde a
\eeq
There is no Chern-Simons term for $a$, which means the fermionic matter field that carries a fundamental representation of the $SU(2)$ gauge field $a$ is confined. The possible elementary deconfined topological excitations should carry charge-1 under $\beta$ or charge-2 under $\tilde a$. Using the K-matrix formalism it is easy to verify that these excitation are all bosons and they have no mutual braiding. Therefore, the resulting state actually contains no nontrivial anyon. Furthermore, one can check that integrating out the gauge fields in Eq. (\ref{eq: positive m, Nf=2}) generates no gravitational Chern-Simons term, which means that this theory has a zero chiral central charge on its edge.

Therefore, the case with $m\gg 1$ and $N_f=2$ is also a short-range entangled state, \ie it has no nontrivial topological excitation or nontrivial edge mode.

\section{Parton mean field of the fermionic $U(2)$ critical theory}
\label{app:U(2)-parton_fermionic}

In this appendix, we discuss the parton mean-field ansatz for the ITO and its confinement transitions.
As discussed in the main text, the $U(2)$ parton construction is $\widetilde S^+ = \phi^\dag f_a^\dag f_b^\dag$.
There is a $U(2)$ gauge redundancy, and the $(f_a, f_b)$ is the $U(2)$ fundamental.
We further rewrite $\phi^\dag = c_1^\dag c_2^\dag$.
The mean field Hamiltonian of the fermionic partons ($c, f$) generally has the first, second and third nearest-neighbor hoppings, which should be consistent with the symmetries: translation symmetry, inversion $C_2$ and $\mathcal T \sigma^*$.
We note that the symmetry actions of translation and inversion are simple on $(c, f)$, while the $\mathcal T \sigma^*$ symmetry transformation is implemented as,
\begin{eqnarray}
\mathcal T \sigma^*: \quad  i &\rightarrow& -i, \\
 \widetilde S_{\vec r}^{x,y}  &\rightarrow& \widetilde S_{\sigma \vec r}^{x,y}, \\
  \widetilde S_{\vec r}^{z}  &\rightarrow& -\widetilde S_{\sigma \vec r}^{z}, \\
   c_{\vec r}^\dag  &\rightarrow& c_{\sigma \vec r}, \\ 
   f_{\vec r}^\dag  &\rightarrow& f_{\sigma \vec r}. 
\end{eqnarray}

\begin{figure}[h]
    \centering
    \includegraphics[width=0.25\linewidth]{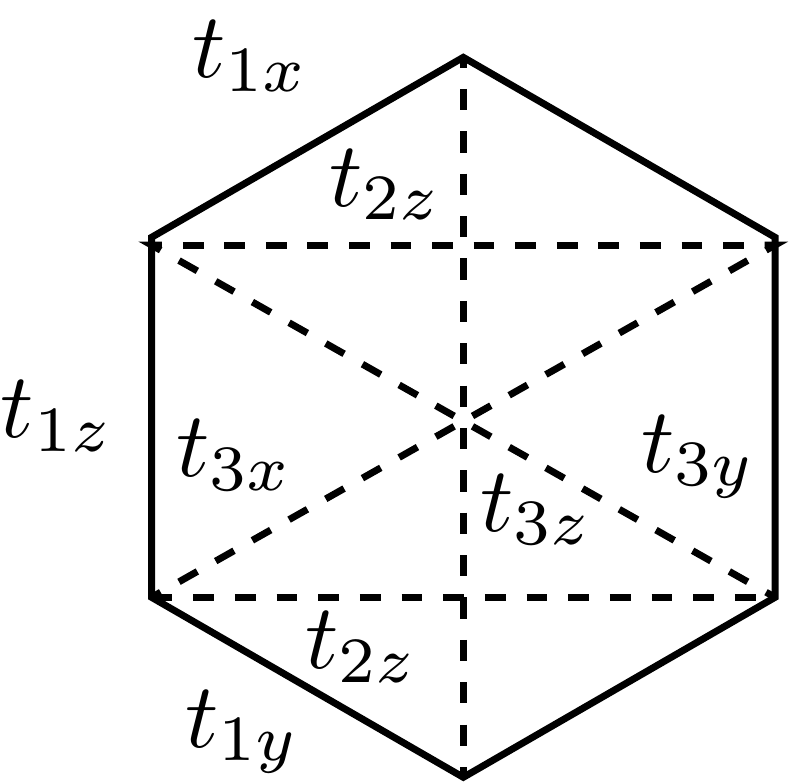}
    \caption{The hopping terms of the parton mean-field ansatz.}
    \label{fig:hopping}
\end{figure}

The mean-field Hamiltonian for the partons (both $c$ and $f$) takes a generic form with the first-, second- and third-nearest-neighbor hoppings, $H=-\sum_{ij} t_{ij} d_i^\dag d_j$, where $d$ can represent either $f$ or $c$.
Specifically, we consider a symmetry preserving hopping pattern, which has parameters $t_{1x}=t_{1y}$, $t_{1z}$, $t_{2z}$, $t_{3x}=t_{3y}$ and $t_{3z}$, as shown in Fig.~\ref{fig:hopping}.

The  parton $c_{1,2}$ is always in a $C=-1$ band, it corresponds to $\phi$ realizing a $\nu=-1/2$ bosonic Laughlin state.
Specifically we take the hopping amplitude as $t^{c_1}_{1x}=t^{c_1}_{1x}=1$, $t^{c_1}_{1z}=1$, $t^{c_1}_{2z}=0.5e^{i\pi/2}$, and $t^{c_2}_{1x}=t^{c_2}_{1x}=1$, $t^{c_2}_{1z}=-1$, $t^{c_2}_{2z}=0.5e^{i\pi/2}$.

The ITO is realized by putting $U(2)$ $f$-partons into a $C=2$ band, which for example can be realized with hopping amplitude  $t_{1x}^f=t_{1y}^f=1$, $t_{1z}^f=1$, $t_{2z}^f= 0.5 e^{i\pi/2}$, $t_{3x}^f=t_{3y}^f=0.3$, $t_{3z}^f=1$.
To realize the zigzag magnetic order, we need to tune the Chern number of $f$-partons to $C=1$.
It can be triggered by tuning $t_{1x}^f=t_{1y}^f$, and the transition happens at $t_{1x}^f=t_{1y}^f=1.3$.
Using this mean-field ansatz, we work out the symmetry quantum numbers of the relevant operators as summarized in Table~\ref{tab:symmetry}.
We note that the quantum numbers of $d\left(\Tr\mb{a}\right)$ turn out to be identical to those of $\bar\Psi\gamma^\mu\Psi$ in all our fermionic dual theories, so they are not displayed in the tables.

\begin{table}
\setlength{\tabcolsep}{0.2cm}
\renewcommand{\arraystretch}{1.4}
    \centering
    \begin{tabular}{c|ccccc} 
    \hline \hline
& $T_{1}$ & $T_2$ & $C_2$ & $\mathcal T \sigma^*$ \\ \hline
$\bar \Psi \tau^{x} \Psi$ & $-1$ & $-1$ & $-1$ & $1$  \\
$\bar \Psi \tau^{y} \Psi$ & $-1$ & $-1$ & $-1$ & $-1$  \\
$\bar \Psi \tau^{z} \Psi$ & $1$ & $1$ & $1$ & $-1$  \\
    
$\bar \Psi \gamma^0 \Psi$ & $1$ & $1$ & $1$ & $-1$ \\ 
$\bar \Psi \gamma^1 \Psi$ & $1$ & $1$ & $-1$ & $1$ \\ 
$\bar \Psi \gamma^2 \Psi$ & $1$ & $1$ & $-1$ & $-1$ \\ 

$\bar \Psi \gamma^0 \tau^{x} \Psi$ & $-1$ & $-1$ & $-1$ & $-1$ \\ 
$\bar \Psi \gamma^1 \tau^{x} \Psi$ & $-1$ & $-1$ & $1$ & $1$ \\ 
$\bar \Psi \gamma^2 \tau^{x} \Psi$ & $-1$ & $-1$ & $1$ & $-1$ \\   

$\bar \Psi \gamma^0 \tau^{y} \Psi$ & $-1$ & $-1$ & $-1$ & $1$ \\ 
$\bar \Psi \gamma^1 \tau^{y} \Psi$ & $-1$ & $-1$ & $1$ & $-1$ \\ 
$\bar \Psi \gamma^2 \tau^{y} \Psi$ & $-1$ & $-1$ & $1$ & $1$ \\   

$\bar \Psi \gamma^0 \tau^{z} \Psi$ & $1$ & $1$ & $1$ & $1$ \\ 
$\bar \Psi \gamma^1 \tau^{z} \Psi$ & $1$ & $1$ & $-1$ & $-1$ \\ 
 $\bar \Psi \gamma^2 \tau^{z} \Psi$ & $1$ & $1$ & $-1$ & $1$ \\ 
  $\mathcal M_1$ & $-1$ & $-1$ & $-1$ & $ (-1)^s\mathcal M_2^\dag $  \\       
    $\mathcal M_2$ & $-1$ & $-1$ & $-1$ & $(-1)^s\mathcal M_1^\dag $  \\       
  $\mathcal M_3$ & $1$ & $1$ & $1$ & $-(-1)^s \mathcal M_3^\dag $  \\   
 \hline\hline
    \end{tabular}
    \caption{Symmetries of operators in the $N_f=2$ critical theory with Dirac nodes at the $M_1$ and $M_2$ points. 
    $s=0,1$ cannot be determined using our current method.
  There is one symmetry allowed relevant operator, $\bar \Psi \gamma^0 \tau^{z} \Psi$, which will destroy the quantum critical point.}
    \label{tab:symmetry_Nf2_M12}
\end{table}

\begin{table}
\setlength{\tabcolsep}{0.2cm}
\renewcommand{\arraystretch}{1.4}
    \centering
    \begin{tabular}{c|ccccc} 
    \hline \hline
    & $T_{1}$ & $T_2$ & $C_2$ & $\mathcal T \sigma^*$ \\ \hline
    $\bar \Psi_1  \Psi_2$ & $e^{-2ik}$ & $e^{-2ik}$ & $-\bar\Psi_2\Psi_1$ & $-\bar\Psi_2\Psi_1$  \\
    $\bar \Psi_2  \Psi_1$ & $e^{2ik}$ & $e^{2ik}$ & $-\bar\Psi_1\Psi_2$ & $-\bar\Psi_1\Psi_2$ \\
    $\bar \Psi \tau^{z} \Psi$ & $1$ & $1$ & $-1$ & $1$  \\
    
$\bar \Psi \gamma^0 \Psi$ & $1$ & $1$ & $1$ & $-1$ \\ 
$\bar \Psi \gamma^1 \Psi$ & $1$ & $1$ & $-1$ & $1$ \\ 
$\bar \Psi \gamma^2 \Psi$ & $1$ & $1$ & $-1$ & $-1$ \\ 

$\bar \Psi_1 \gamma^0 \Psi_2$ & $e^{-2ik}$ & $e^{-2ik}$ & $-\bar\Psi_2\gamma^0\Psi_1$ & $\bar\Psi_2\gamma^0\Psi_1$ \\ 
$\bar \Psi_1 \gamma^1 \Psi_2$ & $e^{-2ik}$ & $e^{-2ik}$ & $\bar\Psi_2\gamma^1\Psi_1$ & $-\bar\Psi_2\gamma^1\Psi_1$ \\ 
$\bar \Psi_1 \gamma^2 \Psi_2$ & $e^{-2ik}$ & $e^{-2ik}$ & $\bar\Psi_2\gamma^2\Psi_1$ & $\bar\Psi_2\gamma^2\Psi_1$ \\

$\bar \Psi_2 \gamma^0 \Psi_1$ & $e^{2ik}$ & $e^{2ik}$ & $-\bar \Psi_1 \gamma^0 \Psi_2$ & $\bar\Psi_1\gamma^0\Psi_2$ \\ 
$\bar \Psi_2 \gamma^1 \Psi_1$ & $e^{2ik}$ & $e^{2ik}$ & $\bar \Psi_1 \gamma^1 \Psi_2$ & $-\bar\Psi_1\gamma^1\Psi_2$ \\ 
$\bar \Psi_2 \gamma^2 \Psi_1$ & $e^{2ik}$ & $e^{2ik}$ & $\bar \Psi_1 \gamma^2 \Psi_2$ & $\bar\Psi_1\gamma^2\Psi_2$ \\

$\bar \Psi \gamma^0 \tau^{z} \Psi$ & $1$ & $1$ & $-1$ & $-1$ \\ 
$\bar \Psi \gamma^1 \tau^{z} \Psi$ & $1$ & $1$ & $1$ & $1$ \\ 
$\bar \Psi \gamma^2 \tau^{z} \Psi$ & $1$ & $1$ & $1$ & $-1$ \\ 

  $\mathcal M_1$ & $-e^{2ik}$ & $-e^{2ik}$ & $-\mathcal M_2$ & $(-1)^s\mc{M}_1^\dag$  \\       
    $\mathcal M_2$ & $-e^{-2ik}$ & $-e^{-2ik}$ & $-\mathcal M_1$ & $(-1)^s\mc{M}_2^\dag$  \\       
  $\mathcal M_3$ & $-1$ & $-1$ & $-1$ & $-(-1)^s \mathcal M_3^\dag $  \\   
  \hline\hline
    \end{tabular}
    \caption{Symmetries of operators in the $N_f=2$ critical theory with two Dirac cones on the high symmetry line $K-K'$. 
    $s=0,1$ cannot be determined using our current method.
    There is one symmetry allowed relevant operator, $\bar\Psi\gamma^1\tau^z\Psi$, which however only moves the location of Dirac points without destroying the quantum critical point. }
    \label{tab:symmetry_Nf2_K}
\end{table}

\begin{table}
\setlength{\tabcolsep}{0.2cm}
\renewcommand{\arraystretch}{1.4}
    \centering
    \begin{tabular}{c|ccccc} 
    \hline \hline
    & $T_{1}$ & $T_2$ & $C_2$ & $\mathcal T \sigma^*$ \\ \hline
    $\bar \Psi_1  \Psi_2$ & $e^{-2ik}$ & $e^{2ik}$ & $-\bar \Psi_2  \Psi_1$ & $1$  \\
    $\bar \Psi_2  \Psi_1$ & $e^{2ik}$ & $e^{-2ik}$ & $-\bar \Psi_1  \Psi_2$ & $1$  \\
    $\bar \Psi \tau^{z} \Psi$ & $1$ & $1$ & $-1$ & $-1$  \\
    
$\bar \Psi \gamma^0 \Psi$ & $1$ & $1$ & $1$ & $-1$ \\ 
$\bar \Psi \gamma^1 \Psi$ & $1$ & $1$ & $-1$ & $1$ \\ 
$\bar \Psi \gamma^2 \Psi$ & $1$ & $1$ & $-1$ & $-1$ \\ 

$\bar \Psi_1 \gamma^0 \Psi_2$ & $e^{-2ik}$ & $e^{2ik}$ & $-\bar \Psi_2 \gamma^0 \Psi_1$ & $-1$ \\ 
$\bar \Psi_1 \gamma^1 \Psi_2$ & $e^{-2ik}$ & $e^{2ik}$ & $\bar \Psi_2 \gamma^1 \Psi_1$ & $1$ \\ 
$\bar \Psi_1 \gamma^2 \Psi_2$ & $e^{-2ik}$ & $e^{2ik}$ & $\bar \Psi_2 \gamma^2 \Psi_1$ & $-1$ \\

$\bar \Psi_2 \gamma^0 \Psi_1$ & $e^{2ik}$ & $e^{-2ik}$ & $-\bar \Psi_1 \gamma^0 \Psi_2$ & $-1$ \\ 
$\bar \Psi_2 \gamma^1 \Psi_1$ & $e^{2ik}$ & $e^{-2ik}$ & $\bar \Psi_1 \gamma^1 \Psi_2$ & $1$ \\ 
$\bar \Psi_2 \gamma^2 \Psi_1$ & $e^{2ik}$ & $e^{-2ik}$ & $\bar \Psi_1 \gamma^2 \Psi_2$ & $-1$ \\

$\bar \Psi \gamma^0 \tau^{z} \Psi$ & $1$ & $1$ & $-1$ & $1$ \\ 
$\bar \Psi \gamma^1 \tau^{z} \Psi$ & $1$ & $1$ & $1$ & $-1$ \\ 
$\bar \Psi \gamma^2 \tau^{z} \Psi$ & $1$ & $1$ & $1$ & $1$ \\ 

  $\mathcal M_1$ & $-e^{2ik}$ & $-e^{-2ik}$ & $-\mathcal M_2$ & $ -(-1)^s\mathcal M_2^\dag $  \\       
    $\mathcal M_2$ & $-e^{-2ik}$ & $-e^{2ik}$ & $-\mathcal M_1$ & $-(-1)^s\mathcal M_1^\dag $  \\       
  $\mathcal M_3$ & $-1$ & $-1$ & $-1$ & $(-1)^s \mathcal M_3^\dag $  \\   
 \hline\hline
    \end{tabular}
    \caption{Symmetries of operators in the $N_f=2$ critical theory with two Dirac cones on the high symmetry line $M_3-M_3$. $s=0,1$ cannot be determined using our current method.
    There is one symmetry allowed relevant operator, $\bar \Psi \gamma^2 \tau^{z} \Psi$, which however only moves the location of Dirac points without destroying the quantum critical point.}
    \label{tab:symmetry_Nf2_M3}
\end{table}

To realize the transition from the ITO to the trivially polarized state, we need to tune the Chern number directly from $C=2$ to $C=0$.
There are three different types of ways to realize this transition:
\begin{enumerate}
    \item Tuninng $t_{1z}^f$, and the transition happens at $t_{1z}^f =1.6$. 
    The two Dirac cones are at the $M_1$ and $M_2$ points. In this case, the symmetry actions on $\Psi$ are given by
    \beq
    \begin{split}
    &T_1: \Psi\rightarrow-\tau^z\Psi\\
    &T_2: \Psi\rightarrow\tau^z\Psi\\
    &C_2: \Psi(\vec r)\rightarrow\gamma^0\tau^z\Psi(-\vec r)\\
    &\mc{T}\sigma^*: \Psi(x, y)\rightarrow i\gamma^1\tau^x\Psi^\dag(x, -y)
    \end{split}
    \eeq
    
    The quantum numbers of the gauge invariant relevant operators are summarized in Table~\ref{tab:symmetry_Nf2_M12}.

    \item Tunning $t_{3x}^f=t_{3y}^f$, and the transition happens at $t_{3x}^f=t_{3y}^f=1$. 
    The two Dirac cones are at $(k_1, k_2)=(k, k), (-k, -k)$, which are on the high symmetry line $K-K'$. 
    In this case, the symmetry actions on $\Psi$ are given by
    \beq
    \begin{split}
    &T_1: \Psi\rightarrow e^{ik\tau^z}\Psi\\
    &T_2: \Psi\rightarrow e^{ik\tau^z}\Psi\\
    &C_2: \Psi(\vec r)\rightarrow \gamma^0\tau^y\Psi(-\vec r)\\
    &\mc{T}\sigma^*: \Psi(x, y)\rightarrow  i\gamma^1\tau^z\Psi^\dag(x, -y)
    \end{split}
    \eeq
    
    The quantum numbers of the gauge invariant relevant operators are summarized in Table~\ref{tab:symmetry_Nf2_K}.   
    \item Tuning $t_{3z}^f$, and the transition happens at $t_{3z}^f=1.6$.
    The two Dirac cones are at $(k_1, k_2)=(k, -k), (-k, k)$, which are on the high symmetry line $M_3-M_3$. 
    In this case, the symmetry actions on $\Psi$ are given by
    \beq
    \begin{split}
    &T_1: \Psi\rightarrow e^{ik\tau^z}\Psi\\
    &T_2: \Psi\rightarrow e^{-ik\tau^z}\Psi\\
    &C_2: \Psi(\vec r)\rightarrow \gamma^0\tau^y\Psi(-\vec r)\\
    &\mc{T}\sigma^*: \Psi(x, y)\rightarrow i\gamma^1\tau^x\Psi^\dag(x, -y)
    \end{split}
    \eeq
    The quantum numbers of the gauge invariant relevant operators are summarized in Table~\ref{tab:symmetry_Nf2_M3}. 
\end{enumerate}

Finally, we make a few comments on the monopole operators. 
Technically, we follow the method 
in Ref.~\cite{Kapustin2002,Alicea_monopole, Song2018a, Song2018b} to calculate the quantum numbers of the monopoles.
Namely, we explicitly construct the monopole states on a torus, and then extract the quantum number of the monopole states.
Specifically, we put the system on a $2\times L \times L$ lattice, and spread a uniform $2\pi$ flux for each parton $c, f$.
Each Dirac fermion will  form Landau levels with one exact zero mode.
When $N_f=1$, the gauge invariant monopole corresponds to a state with all negative-energy Fermi sea filled.
In contrast, in the $N_f=2$ critical theory, the gauge invariant monopoles should have two zero modes filled (each from one $U(2)$ color) together with the filled negative-energy Fermi sea.
There are three gauge invariant ways to fill the zero modes, 
\begin{equation} \label{eq: monopole operators}
\mathcal M_1=\widetilde{\mathcal{M}}\psi_{1a}\psi_{1b}, \quad 
\mathcal M_2=\widetilde{\mathcal{M}}\psi_{2b}\psi_{2a}, \quad
\mathcal M_3=\frac{1}{\sqrt{2}}\widetilde{\mathcal{M}}(\psi_{1a}\psi_{2b}-\psi_{1b}\psi_{2a}).
\end{equation}
Here $\widetilde{\mathcal M}$ is the bare monopole with $2\pi$ flux and filled negative-energy Fermi sea.
$\psi$ represents the zero mode, and $1,2$ are the flavor indices and $a,b$ are the color indices. 
The three monopoles are in the adjoint representation of the $SU(2)$ flavor symmetry.
Using our current method, we are not able to determine the quantum number of the monopoles under $\mathcal T \sigma^*$, for which there is an undetermined sign $\mathcal M\rightarrow \pm \mathcal M^\dag $.
In the $N_f=1$ theory, we speculate the sign is $-1$, hence it matches the quantum number of the zigzag order.
In the $N_f=2$ theory, we leave this sign undetermined, and it has no influence on our discussion on the nature of the confined state.

\subsection*{More details on the quantum numbers of monopoles}

Before finishing this appendix, we discuss in more details the symmetry actions on the monopoles of the fermionic critical theory with $N_f=2$, using state-operator correspondence. Including both colors and spins, there will be four zero modes in the presence of $\pm 2\pi$ background flux. In this case there is no Chern-Simons term for $\Tr(\mb{a})$, so two of the zero modes need to be occupied to form a gauge invariant state (operator). In terms of states, there are three different ways to occupy these zero modes and make a color singlet:
\beq \label{eq: states}
f_{1a}^\dag f_{1b}^\dag|0\ra,
\quad
f_{2b}^\dag f_{2a}^\dag|0\ra,
\quad
\frac{1}{\sqrt{2}}(f_{1a}^\dag f_{2b}^\dag-f_{1b}^\dag f_{2a}^\dag)|0\ra
\eeq
where the $f$'s are the operators of the zero modes, and $|0\ra$ is the ground state under a $2\pi$ background flux with no zero mode occupied. We use $1$ and $2$ to label the two different flavors, and $a$ and $b$ to label the two different colors. These states correspond to the operators $\mc{M}_{1,2,3}$ in Eq (\ref{eq: monopole operators}), respectively. 

The actions of $T_{1,2}$ and $C_2$ are simpler because they do not take the monopole operators to their Hermitian conjugates. To determine the action of $\mc{T}\sigma^*$, which takes the monopoles to their Hermitian conjugates, it will be important to first identify the corresponding states of the Hermitian conjugates of these operators. This can be worked out using the methods in Refs. \cite{Zou2018, Zou2018a}. More precisely, let us write the three states in Eq. (\ref{eq: states}) in a more suggestive form
\beq
\begin{split}
&\mc{M}_1\sim f_{1a}^\dag f_{1b}^\dag|0\ra=\left(f^T\frac{(1+\tau^z)\epsilon}{4}f\right)^*|0\ra =\left(f^T\tau^y\frac{(\tau^y+i\tau^x)\epsilon}{4} f\right)^*|0\ra\\
&\mc{M}_2\sim f_{2b}^\dag f_{2a}^\dag|0\ra=\left(-f^T\frac{(1-\tau^z)\epsilon}{4}f\right)^*|0\ra =\left(-f^T\tau^y\frac{(\tau^y-i\tau^x)\epsilon}{4}f\right)^*|0\ra\\
&\mc{M}_3\sim\frac{1}{\sqrt{2}}(f_{1a}^\dag f_{2b}^\dag-f_{1b}^\dag f_{2a}^\dag)|0\ra=\frac{1}{\sqrt{2}}\left(f^T\frac{\tau^x\epsilon}{2}f\right)^*|0\ra =\frac{1}{\sqrt{2}}\left(f^T\tau^y\frac{-i\tau^z\epsilon}{2}f\right)^*|0\ra
\end{split}
\eeq
where $\tau$ acts on the flavor space and $\epsilon$ acts on the color space. From these we get
\beq
\begin{split}
&i(\mc{M}_1+\mc{M}_2)\sim \left(f^T\tau^y\tau^x\frac{\epsilon}{2}f\right)^*|0\ra\\
&\mc{M}_1-\mc{M}_2\sim \left(f^T\tau^y\tau^y\frac{\epsilon}{2}f\right)^*|0\ra\\
&-i\mc{M}_3\sim \frac{1}{\sqrt{2}}\left(f^T\tau^y\tau^z\frac{\epsilon}{2}f\right)^*|0\ra
\end{split}
\eeq

Therefore, $(i(\mc{M}_1+\mc{M}_2), \mc{M}_1-\mc{M}_2, -i\mc{M}_3)$ transforms as a vector under the $SU(2)$ flavor symmetry. Because this representation of the $SU(2)$ transformation is real, $(-i(\mc{M}_1^\dag+\mc{M}_2^\dag), \mc{M}_1^\dag-\mc{M}_2^\dag, i\mc{M}_3^\dag)$ also transforms in the same representation under the $SU(2)$ flavor symmetry. This observation tells us what the corresponding states of these Hermitian conjugates are (up to an undetermined phase factor): 
\beq
\begin{split}
&-i(\mc{M}_1^\dag+\mc{M}_2^\dag)\sim
\left(\tilde f^T\tau^y\tau^x\frac{\epsilon}{2}\tilde f\right)^*|\tilde 0\ra\\
&\mc{M}_1^\dag-\mc{M}_2^\dag\sim \left(\tilde f^T\tau^y\tau^y\frac{\epsilon}{2}\tilde f\right)^*|\tilde 0\ra\\
&i\mc{M}_3^\dag\sim
\frac{1}{\sqrt{2}}\left(\tilde f^T\tau^y\frac{\tau^z\epsilon}{2}\tilde f\right)^*|\tilde 0\ra
\end{split}
\eeq
where $|\tilde 0\ra$ is the ground state under a $-2\pi$ background flux with no zero modes occupied, and $\tilde f$'s are the corresponding zero modes under a $-2\pi$ flux background. From the above we get
\beq
\begin{split}
&\mc{M}^\dag_1\sim-\tilde f_{2b}^\dag\tilde f_{2a}^\dag|\tilde 0\ra\\
&\mc{M}^\dag_2\sim -\tilde f_{1a}^\dag\tilde f_{1b}^\dag|\tilde 0\ra\\
&\mc{M}^\dag_3\sim-\frac{1}{\sqrt{2}}\left(\tilde f_{1a}^\dag\tilde f_{2b}^\dag-\tilde f_{1b}^\dag\tilde f_{2a}^\dag\right)|\tilde 0\ra
\end{split}
\eeq

Now we can check the action of $\mc{T}\sigma^*$ on $\mc{M}_{1,2,3}$. We have two types of actions of $\mc{T}\sigma^*$ on the fermions. For the first type:
\beq
\mc{T}\sigma^*: \Psi(x, y)\rightarrow i\gamma^1\tau^x\Psi(x, -y)^\dag
\eeq
we have
\beq
\begin{split}
&\mc{M}_1\sim f^\dag_{1a}f_{1b}^\dag|0\ra\rightarrow \tilde f_{2a}\tilde f_{2b}\tilde f_{1a}^\dag\tilde f_{1b}^\dag \tilde f_{2a}^\dag\tilde f_{2b}^\dag|\tilde 0\ra=-\tilde f_{1a}^\dag\tilde f_{1b}^\dag|\tilde 0\ra\sim\mc{M}_2^\dag\\
&\mc{M}_2\sim f^\dag_{2b}f^\dag_{2a}|0\ra\rightarrow \tilde f_{1b}\tilde f_{1a}\tilde f_{1a}^\dag\tilde f_{1b}^\dag \tilde f_{2a}^\dag\tilde f_{2b}^\dag|\tilde 0\ra=\tilde f_{2a}^\dag\tilde f_{2b}^\dag|\tilde 0\ra\sim\mc{M}_1^\dag\\
&\mc{M}_3\sim\frac{1}{\sqrt{2}}\left(f^\dag_{1a}f^\dag_{2b} -f^\dag_{1b}f^\dag_{2a}\right)|0\ra\rightarrow\frac{1}{\sqrt{2}} \left(\tilde f_{2a}\tilde f_{1b}-\tilde f_{2b}\tilde f_{1a}\right)\tilde f_{1a}^\dag\tilde f_{1b}^\dag \tilde f_{2a}^\dag\tilde f_{2b}^\dag|\tilde 0\ra\\
&\qquad\qquad\qquad\qquad\qquad\qquad\qquad
=\frac{1}{\sqrt{2}}\left(\tilde f^\dag_{1a}\tilde f^\dag_{2b}-\tilde f_{1b}^\dag \tilde f_{2a}^\dag\right)|\tilde 0\ra\sim-\mc{M}_3^\dag
\end{split}
\eeq
In the above we have taken the convention that, under $\mc{T}\sigma ^*$, $|0\ra\rightarrow\tilde f_{1a}^\dag\tilde f_{1b}^\dag\tilde f_{2a}^\dag\tilde f_{2b}^\dag|\tilde 0\ra$. Notice these transformation rules have a common undetermined phase factor for $\mc{M}_{1,2,3}$.

For the second type of $\mc{T}\sigma^*$:
\beq
\mc{T}\sigma^*: \Psi(x, y)\rightarrow i\gamma^1\tau^z\Psi(x, -y)^\dag
\eeq
we have
\beq
\begin{split}
&\mc{M}_1\sim f^\dag_{1a}f_{1b}^\dag|0\ra\rightarrow \tilde f_{1a}\tilde f_{1b}\tilde f_{1a}^\dag\tilde f_{1b}^\dag \tilde f_{2a}^\dag\tilde f_{2b}^\dag|\tilde 0\ra=-\tilde f_{2a}^\dag\tilde f_{2b}^\dag|\tilde 0\ra\sim-\mc{M}_1^\dag\\
&\mc{M}_2\sim f^\dag_{2b}f^\dag_{2a}|0\ra\rightarrow \tilde f_{2b}\tilde f_{2a}\tilde f_{1a}^\dag\tilde f_{1b}^\dag \tilde f_{2a}^\dag\tilde f_{2b}^\dag|\tilde 0\ra=\tilde f_{1a}^\dag\tilde f_{1b}^\dag|\tilde 0\ra\sim-\mc{M}_2^\dag\\
&\mc{M}_3\sim\frac{1}{\sqrt{2}}\left(f^\dag_{1a}f^\dag_{2b} -f^\dag_{1b}f^\dag_{2a}\right)|0\ra\rightarrow\frac{1}{\sqrt{2}} \left(-\tilde f_{1a}\tilde f_{2b}+\tilde f_{1b}\tilde f_{2a}\right)\tilde f_{1a}^\dag\tilde f_{1b}^\dag \tilde f_{2a}^\dag\tilde f_{2b}^\dag|\tilde 0\ra\\
&\qquad\qquad\qquad\qquad\qquad\qquad\qquad
=\frac{1}{\sqrt{2}}\left(\tilde f^\dag_{1b}\tilde f^\dag_{2a}-\tilde f_{1a}^\dag \tilde f_{2b}^\dag\right)|\tilde 0\ra\sim\mc{M}_3^\dag
\end{split}
\eeq
Again, there is a common undetermined phase factor for the transformation rules of $\mc{M}_{1,2,3}$.

To summarize, this discussion tells us about the action of $\mc{T}\sigma^*$ on the monopole operators in the fermionic critical theory with $N_f=2$.
Similar arguments can establish the actions of other symmetries on these monopoles. 
However, this method leaves an undetermined common phase factor in each transformation of $\mc{M}_{1,2,3}$. 
For unitary symmetries $T_{1,2}$ and $C_2$, we determine this phase factor numerically. 
For the anti-unitary symmetry $\mc{T}\sigma^*$, the current numerical method is insufficient to pin down this phase factor, and we leave it open. 
The results are listed in Tables \ref{tab:symmetry_Nf2_M12} \ref{tab:symmetry_Nf2_K} and \ref{tab:symmetry_Nf2_M3}. 
Notice for the fermionic critical theory with $N_f=2$, the quantum numbers of monopoles will not affect the nature of the confined phase at all. 
Furthermore, in the cases with the Dirac points located at two generic momenta on the $K-K'$ line or the $M_3-M_3$ line, where no symmetry allowed fermion bilinear perturbation can destroy the critical point, the unitary symmetries already forbid single monopole operators, while two-fold monopole operators are always symmetry allowed, regardless of what the undetermined phase factors in the actions of $\mc{T}\sigma^*$ are.

\section{Neutron scattering signals at the QCD$_3$-Chern-Simons quantum critical points} \label{app: neutron scattering}

In this appendix we present the details of the analysis of the neutron scattering signals at the QCD$_3$-Chern-Simons critical points, where the main results are summarized in Sec. \ref{sec: exp}.

Because neutron scattering probes the structure factors of the single-spin operators, \eg $\la\widetilde S_i(\vec k)\widetilde S_j(-\vec k)\ra$ with $i,j=x,y,z$, to understand the behavior of these structure factors at those critical points, we need to know which operators in the critical field theories (IR operators) have finite overlap with $\widetilde S_i$. Operators with the same quantum numbers under the global symmetries generically have finite overlap,{\footnote{More precisely, microscopic operators should allow a representation in terms of a summation of IR operators that have the same properties, such as symmetry quantum numbers. So these IR operators and the original microscopic operators generically have finite overlap.}} so we just need to compare the symmetry quantum numbers of $\widetilde S_i$ with those listed in Tables \ref{tab:symmetry_Nf1}, \ref{tab:symmetry_Nf2_K} and \ref{tab:symmetry_Nf2_M3} to determine which of them have overlap. On the other hand, some operators may have a scaling dimension larger than $3/2$. Even if these operators have finite overlap with $\widetilde S_i$, they will show as dips rather than peaks in the neutron scattering spectrum, and we do not consider them because their signals are practically weak. Notice all operators corresponding to conserved currents (\ie fermion bilinears involving $\gamma_\mu$ and $d(\Tr\mathbf{a})$ in Tables \ref{tab:symmetry_Nf1}, \ref{tab:symmetry_Nf2_K} and \ref{tab:symmetry_Nf2_M3}) are of this type. We will also assume all other operators in these tables have scaling dimension smaller than $3/2$, so they will show as peaks in the neutron scattering signals.

\begin{table}
\setlength{\tabcolsep}{0.3cm}
\renewcommand{\arraystretch}{1.4}
    \centering
    \begin{tabular}{c|cccc} 
    \hline \hline
    & $T_{1,2}$ & $C_2$ & $\mathcal T \sigma^*$ \\ \hline
  $\widetilde S_{xA}(\vec r)$  & $\widetilde S_{xA}(\vec r+\vec n_{1,2})$ & $\widetilde S_{xB}(\vec r')$ & $\widetilde S_{xB}(\vec r'')$ \\       
$\widetilde S_{yA}(\vec r)$ & $\widetilde S_{yA}(\vec r+\vec n_{1,2})$ & $\widetilde S_{yB}(\vec r')$ & $\widetilde S_{yB}(\vec r'')$ \\ 
$\widetilde S_{zA}(\vec r)$ & $\widetilde S_{zA}(\vec r+\vec n_{1,2})$ & $\widetilde S_{zB}(\vec r')$ & $-\widetilde S_{zB}(\vec r'')$ \\ 
$\widetilde S_{xB}(\vec r)$ & $\widetilde S_{xB}(\vec r+\vec n_{1,2})$ & $\widetilde S_{xA}(\vec r')$ & $\widetilde S_{xA}(\vec r'')$ \\
$\widetilde S_{yB}(\vec r)$ & $\widetilde S_{yB}(\vec r+\vec n_{1,2})$ & $\widetilde S_{yA}(\vec r')$ & $\widetilde S_{yB}(\vec r'')$ \\ 
$\widetilde S_{zB}(\vec r)$ & $\widetilde S_{zB}(\vec r+\vec n_{1,2})$ & $\widetilde S_{zA}(\vec r')$ & $-\widetilde S_{zB}(\vec r'')$ \\ 
\hline\hline
    \end{tabular}
    \caption{Symmetry transformations of operator $\widetilde S_{ia}(\vec r)$, where $i=x, y, z$ labels the orientation of the spin, $a=A,B$ labels the sublattice, $\vec r$ labels the position of the unit cell, $\vec r'$ is the $C_2$-partner of $\vec r$, and $\vec r''$ is the $\sigma$-partner of $\vec r$.}
    \label{tab:symmetry_spin}
\end{table}

The transformation rules of $\tilde S_i$ under global symmetries are listed in Table \ref{tab:symmetry_spin}. First, we note that at all these three critical points, the singlet mass operator $\bar\Psi\Psi$ has trivial quantum numbers under all symmetries, just as $\widetilde S_{iA}(\Gamma)+\widetilde S_{iB}(\Gamma)$ for $i=x, y$. This will potentially give rise to a strong signal at the $\Gamma$ point in the BZ. Below we specify to other potential strong neutron scattering signals at these critical points.

Let us first analyze the neutron scattering signals at the transition between the zigzag phase and the ITO phase by checking the quantum numbers of operators in Table \ref{tab:symmetry_Nf1}. In this case we only need to consider the monopole operator, $\mathcal{M}$, which turns out to have identical quantum numbers as operators
$$
\widetilde S_{xA}(M_3)-\widetilde S_{xB}(M_3)
{\rm\ and\ }
\widetilde S_{yA}(M_3)-\widetilde S_{yB}(M_3).
$$

These are just the order parameter of the zigzag phase and can be detected by neutron scattering at $M_3$ of the BZ. Furthermore, in polarized neutron scattering experiments, the critical exponent characterizing how fast the signals diverges upon approaching $M_3$ is the same if the spin polarization is along any direction spanned by $\widetilde S_x$ and $\widetilde S_y$, reflecting an emergent $U(1)$ spin rotational symmetry with respect to $\widetilde S_z$ in our theory. For notational simplicity, the above relation of operators with identical quantum numbers will be written in the form
\beq
\mathcal{M}\sim\widetilde S_{iA}(M_3)-\widetilde S_{iB}(M_3)
\eeq
with $i=x, y$.

Next, let us turn to the transition between the ITO phase and the trivial phase, which has two possible symmetry implementations at the critical point, corresponding to Tables \ref{tab:symmetry_Nf2_K} and \ref{tab:symmetry_Nf2_M3}.{\footnote{In these tables we will always consider the case where $k$ is a generic value of momentum.}} In Table \ref{tab:symmetry_Nf2_K}, the operators to be considered include $\bar\Psi_1\Psi_2$, $\bar\Psi_2\Psi_1$, $\bar\Psi\tau^z\Psi$ and $\mathcal{M}_{1,2,3}$. By comparing their symmetry properties with those in Table \ref{tab:symmetry_spin}, we get
\beq \label{eq: symmetry match K}
\begin{split}
&\bar\Psi\tau^z\Psi
\sim
\widetilde S_{zA}(\Gamma)-\widetilde S_{zB}(\Gamma)\\
&
\left(
\begin{array}{c}
\bar\Psi_1\Psi_2\\
-\bar\Psi_2\Psi_1
\end{array}
\right)
\sim
\left(
\begin{array}{c}
\widetilde S_{iA}(-2k, -2k)-\widetilde S_{iB}(-2k, -2k)\\
\widetilde S_{iB}(2k, 2k)-\widetilde S_{iA}(2k, 2k)
\end{array}
\right)
\sim
\left(
\begin{array}{c}
i(\widetilde S_{zA}(-2k, -2k)+\widetilde S_{zB}(-2k, -2k))\\
i(\widetilde S_{zB}(2k, 2k)+\widetilde S_{zA}(2k, 2k))
\end{array}
\right)
\end{split}
\eeq
with $i=x, y$. The identification of $\mathcal{M}_{1,2,3}$ depends on the value of $s$ in Table \ref{tab:symmetry_Nf2_K}. If $s=0$,
\beq
\begin{split}
&\mathcal{M}_3\sim 
\widetilde S_{iA}(M_3)-\widetilde S_{iB}(M_3)\\
&\left(
\begin{array}{c}
\mathcal{M}_1-\mathcal{M}_2^\dag\\
-\mathcal{M}_2+\mathcal{M}_1^\dag
\end{array}
\right)
\sim
\left(
\begin{array}{c}
\widetilde S_{iA}(2k+\pi, 2k+\pi)+\widetilde S_{iB}(2k+\pi, 2k+\pi)\\
\widetilde S_{iB}(-2k+\pi, -2k+\pi)+\widetilde S_{iA}(-2k+\pi, -2k+\pi)
\end{array}
\right)\\
&\qquad\qquad\qquad\quad\ 
\sim
\left(
\begin{array}{c}
i(\widetilde S_{zA}(2k+\pi, 2k+\pi)-\widetilde S_{zB}(2k+\pi, 2k+\pi))\\
i(\widetilde S_{zB}(-2k+\pi, -2k+\pi)-\widetilde S_{zA}(-2k+\pi, -2k+\pi))
\end{array}
\right)\\
&\left(
\begin{array}{c}
\mathcal{M}_1+\mathcal{M}_2^\dag\\
-\mathcal{M}_2-\mathcal{M}_1^\dag
\end{array}
\right)
\sim
\left(
\begin{array}{c}
i(\widetilde S_{iA}(2k+\pi, 2k+\pi)+\widetilde S_{iB}(2k+\pi, 2k+\pi))\\
i(\widetilde S_{iB}(-2k+\pi, -2k+\pi)+\widetilde S_{iA}(-2k+\pi, -2k+\pi))
\end{array}
\right)\\
&\qquad\qquad\qquad\quad\ 
\sim
\left(
\begin{array}{c}
\widetilde S_{zA}(2k+\pi, 2k+\pi)-\widetilde S_{zB}(2k+\pi, 2k+\pi)\\
\widetilde S_{zB}(-2k+\pi, -2k+\pi)-\widetilde S_{zA}(-2k+\pi, -2k+\pi)
\end{array}
\right)
\end{split}
\eeq
with $i=x, y$. If $s=1$,
\beq
\begin{split}
    &\mathcal{M}_3\sim\widetilde S_{zA}(M_3)-\widetilde S_{zB}(M_3)\\
    &
    \left(
    \begin{array}{c}
    \mathcal{M}_1-\mathcal{M}_2^\dag\\
    -\mathcal{M}_2+\mathcal{M}_1^\dag
    \end{array}
    \right)
    \sim
    \left(
    \begin{array}{c}
    i(\widetilde S_{iA}(2k+\pi, 2k+\pi)-\widetilde S_{iB}(2k+\pi, 2k+\pi))\\
    i(\widetilde S_{iB}(-2k+\pi, -2k+\pi)-\widetilde S_{iA}(-2k+\pi, -2k+\pi))
    \end{array}
    \right)\\
    &\qquad\qquad\qquad\quad\ 
    \sim
    \left(
    \begin{array}{c}
    \widetilde S_{zA}(2k+\pi, 2k+\pi)+\widetilde S_{zB}(2k+\pi, 2k+\pi)\\
    \widetilde S_{zB}(-2k+\pi, -2k+\pi)+\widetilde S_{zA}(-2k+\pi, -2k+\pi)
    \end{array}
    \right)\\
    &\left(
    \begin{array}{c}
        \mathcal{M}_1+\mathcal{M}_2^\dag\\
        -\mathcal{M}_2-\mathcal{M}_1^\dag
    \end{array}
    \right)
    \sim
    \left(
    \begin{array}{c}
    \widetilde S_{iA}(2k+\pi, 2k+\pi)-\widetilde S_{iB}(2k+\pi, 2k+\pi)\\
    \widetilde S_{iB}(-2k+\pi, -2k+\pi)-\widetilde S_{iA}(-2k+\pi, -2k+\pi)
    \end{array}
    \right)\\
    &\qquad\qquad\qquad\quad\ 
    \sim
    \left(
    \begin{array}{c}
        i(\widetilde S_{zA}(2k+\pi, 2k+\pi)+\widetilde S_{zB}(2k+\pi, 2k+\pi))\\
        i(\widetilde S_{zB}(-2k+\pi, -2k+\pi)+\widetilde S_{zA}(-2k+\pi, -2k+\pi))
    \end{array}
    \right)
\end{split}
\eeq

From this comparison, we see that in this case the neutron scattering may see strong signals at the $\Gamma$, $M_3$, $(\pm 2k, \pm 2k)$ and $(\pm 2k+\pi, \pm 2k+\pi)$ points in the BZ. Also, we can read off which spin polarizations give rise to power-law divergence in the neutron spectrum in each case. In addition, due to an emergent $SO(3)$ flavor symmetry under which $\bar\Psi\vec\tau\Psi$ and $\mathcal{M}_{1,2,3}$ form two vectors, the critical exponent characterizing the divergence at $\Gamma$ is the same as the critical exponent characterizing the divergence at $(\pm 2k, \pm 2k)$, and the critical exponent at $M_3$ is the same as that at $(\pm 2k+\pi, \pm 2k+\pi)$.

Similarly, in Table \ref{tab:symmetry_Nf2_M3}, the operators to be considered are again $\bar\Psi_1\Psi_2$, $\bar\Psi_2\Psi_1$, $\bar\Psi\tau^z\Psi$ and $\mathcal{M}_{1,2,3}$. By comparing their symmetry properties with the operators in Table \ref{tab:symmetry_spin}, we get
\beq \label{eq: symmetry match M3}
\begin{split}
&\bar\Psi\tau^z\Psi
\sim
\widetilde S_{iA}(\Gamma)-\widetilde S_{iB}(\Gamma)\\
&
\left(
\begin{array}{c}
\bar\Psi_1\Psi_2\\
-\bar\Psi_2\Psi_1
\end{array}
\right)
\sim
\left(
\begin{array}{c}
e^{i\alpha}\widetilde S_{zA}(-2k, 2k)-e^{-i\alpha}\widetilde S_{zB}(-2k, 2k)\\
-e^{-i\alpha}\widetilde S_{zA}(2k, -2k)+e^{i\alpha}\widetilde S_{zB}(2k, -2k)
\end{array}
\right)
\end{split}
\eeq
for $\alpha\in\mathbb{R}$. Again, the identification of $\mathcal{M}_{1, 2, 3}$ depends on the value of $s$ in Table \ref{tab:symmetry_Nf2_M3}. If $s=0$,
\beq
\begin{split}
    &\mathcal{M}_3\sim\widetilde S_{zA}(M_3)-\widetilde S_{zB}(M_3)\\
    &
    \left(
    \begin{array}{c}
    \mathcal{M}_1-\mathcal{M}_2^\dag\\
    -\mathcal{M}_2+\mathcal{M}_1^\dag
    \end{array}
    \right)
    \sim
    \left(
    \begin{array}{c}
    e^{i\alpha}\widetilde S_{iA}(2k+\pi, -2k+\pi)+e^{-i\alpha}\widetilde S_{iB}(2k+\pi, -2k+\pi)\\
    e^{-i\alpha}\widetilde S_{iA}(-2k+\pi, 2k+\pi)+e^{i\alpha}\widetilde S_{iB}(-2k+\pi, 2k+\pi)
    \end{array}
    \right)\\
    &\left(
    \begin{array}{c}
    \mathcal{M}_1+\mathcal{M}_2^\dag\\
    -\mathcal{M}_2-\mathcal{M}_1^\dag
    \end{array}
    \right)
    \sim
    \left(
    \begin{array}{c}
    e^{i\alpha}\widetilde S_{iA}(2k+\pi, -2k+\pi)-e^{-i\alpha}\widetilde S_{iB}(2k+\pi, -2k+\pi)\\
    -e^{-i\alpha}\widetilde S_{iA}(-2k+\pi, 2k+\pi)+e^{i\alpha}\widetilde S_{iB}(-2k+\pi, 2k+\pi)
    \end{array}
    \right)
\end{split}
\eeq
with $i=x, y$, for $\alpha\in\mathbb{R}$. If $s=1$,
\beq
\begin{split}
    &\mathcal{M}_3\sim\widetilde S_{iA}(M_3)-\widetilde S_{iB}(M_3)\\
    &
    \left(
    \begin{array}{c}
    \mathcal{M}_1-\mathcal{M}_2^\dag\\
    -\mathcal{M}_2+\mathcal{M}_1^\dag
    \end{array}
    \right)
    \sim
    \left(
    \begin{array}{c}
    e^{i\alpha}\widetilde S_{zA}(2k+\pi, -2k+\pi)+e^{-i\alpha}\widetilde S_{zB}(2k+\pi, -2k+\pi)\\
    e^{-i\alpha}\widetilde S_{zA}(-2k+\pi, 2k+\pi)+e^{i\alpha}\widetilde S_{zB}(-2k+\pi, 2k+\pi)
    \end{array}
    \right)\\
    &
    \left(
    \begin{array}{c}
    \mathcal{M}_1+\mathcal{M}_2^\dag\\
    -\mathcal{M}_2-\mathcal{M}_1^\dag
    \end{array}
    \right)
    \sim
    \left(
    \begin{array}{c}
    e^{i\alpha}\widetilde S_{zA}(2k+\pi, -2k+\pi)-e^{-i\alpha}\widetilde S_{zB}(2k+\pi, -2k+\pi)\\
    -e^{-i\alpha}\widetilde S_{zA}(-2k+\pi, 2k+\pi)+e^{i\alpha}\widetilde S_{zB}(-2k+\pi, 2k+\pi)
    \end{array}
    \right)
\end{split}
\eeq
with $i=x, y$, for $\alpha\in\mathbb{R}$.

From this comparison, we see that in this case the neutron scattering may see strong signals at the $\Gamma$, $M_3$, $(\pm 2k, \mp 2k)$ and $(\pm 2k+\pi, \mp 2k+\pi)$ points in the BZ. Also, we can read off which spin polarizations give rise to power-law divergence in the neutron spectrum in each case. In addition, due to an emergent $SO(3)$ flavor symmetry under which $\bar\Psi\vec\tau\Psi$ and $\mathcal{M}_{1,2,3}$ form two vectors, the critical exponent characterizing the divergence at $\Gamma$ is the same as the critical exponent characterizing the divergence at $(\pm 2k, \mp 2k)$, and the critical exponent at $M_3$ is the same as that at $(\pm 2k+\pi, \mp 2k+\pi)$.

\end{document}